\documentclass[twocolumn,letterpaper,aps,prc,longbibliography,superscriptaddress,nofootinbib,floatfix]{revtex4-2}

\usepackage{amsmath}	
\usepackage{graphicx}	
\usepackage{xspace}	

\newcommand{\pt}{\mbox{$p_T$}\xspace}

\newcommand{\sqs}{\mbox{$\sqrt{s}$}\xspace}
\newcommand{\snn}{\mbox{$\sqrt{s_{_{NN}}}$}\xspace}
\newcommand{\sqsn}{\mbox{$\sqrt{s_{_{NN}}}$}\xspace}

\newcommand{\pp}{\mbox{$p$$+$$p$}\xspace}
\newcommand{\dau}{\mbox{$d$$+$Au}\xspace}

\newcommand{\pau}{\mbox{$p$$+$Au}\xspace}

\newcommand{\la}{\langle}
\newcommand{\ra}{\rangle}

\newcommand{\mean}[1]{\la #1 \ra}

\newcommand{\heau}{\mbox{$^3$He$+$Au}\xspace}

\newcommand{\pa}{\mbox{$p$$+$$A$}\xspace}

\newcommand{\ttpc}{\mbox{3$\times$2PC}\xspace}

\begin{document}


\title{Kinematic dependence of azimuthal anisotropies in 
$p$$+$Au, $d$$+$Au, $^3$He+Au at $\sqrt{s_{_{NN}}}$ = 200 GeV}

\newcommand{\abilene}{Abilene Christian University, Abilene, Texas 79699, USA}
\newcommand{\augie}{Department of Physics, Augustana University, Sioux Falls, South Dakota 57197, USA}
\newcommand{\banaras}{Department of Physics, Banaras Hindu University, Varanasi 221005, India}
\newcommand{\barc}{Bhabha Atomic Research Centre, Bombay 400 085, India}
\newcommand{\baruch}{Baruch College, City University of New York, New York, New York, 10010 USA}
\newcommand{\bnlcoll}{Collider-Accelerator Department, Brookhaven National Laboratory, Upton, New York 11973-5000, USA}
\newcommand{\bnlphys}{Physics Department, Brookhaven National Laboratory, Upton, New York 11973-5000, USA}
\newcommand{\caucr}{University of California-Riverside, Riverside, California 92521, USA}
\newcommand{\charlesczech}{Charles University, Ovocn\'{y} trh 5, Praha 1, 116 36, Prague, Czech Republic}
\newcommand{\ciae}{Science and Technology on Nuclear Data Laboratory, China Institute of Atomic Energy, Beijing 102413, People's Republic of China}
\newcommand{\cns}{Center for Nuclear Study, Graduate School of Science, University of Tokyo, 7-3-1 Hongo, Bunkyo, Tokyo 113-0033, Japan}
\newcommand{\colorado}{University of Colorado, Boulder, Colorado 80309, USA}
\newcommand{\columbia}{Columbia University, New York, New York 10027 and Nevis Laboratories, Irvington, New York 10533, USA}
\newcommand{\czechtech}{Czech Technical University, Zikova 4, 166 36 Prague 6, Czech Republic}
\newcommand{\debrecen}{Debrecen University, H-4010 Debrecen, Egyetem t{\'e}r 1, Hungary}
\newcommand{\elte}{ELTE, E{\"o}tv{\"o}s Lor{\'a}nd University, H-1117 Budapest, P{\'a}zm{\'a}ny P.~s.~1/A, Hungary}
\newcommand{\eszterhazy}{Eszterh\'azy K\'aroly University, K\'aroly R\'obert Campus, H-3200 Gy\"ongy\"os, M\'atrai \'ut 36, Hungary}
\newcommand{\ewha}{Ewha Womans University, Seoul 120-750, Korea}
\newcommand{\famu}{Florida A\&M University, Tallahassee, FL 32307, USA}
\newcommand{\fsu}{Florida State University, Tallahassee, Florida 32306, USA}
\newcommand{\gsu}{Georgia State University, Atlanta, Georgia 30303, USA}
\newcommand{\hiroshima}{Hiroshima University, Kagamiyama, Higashi-Hiroshima 739-8526, Japan}
\newcommand{\howard}{Department of Physics and Astronomy, Howard University, Washington, DC 20059, USA}
\newcommand{\ihepprot}{IHEP Protvino, State Research Center of Russian Federation, Institute for High Energy Physics, Protvino, 142281, Russia}
\newcommand{\illuiuc}{University of Illinois at Urbana-Champaign, Urbana, Illinois 61801, USA}
\newcommand{\inrras}{Institute for Nuclear Research of the Russian Academy of Sciences, prospekt 60-letiya Oktyabrya 7a, Moscow 117312, Russia}
\newcommand{\instpasczech}{Institute of Physics, Academy of Sciences of the Czech Republic, Na Slovance 2, 182 21 Prague 8, Czech Republic}
\newcommand{\isu}{Iowa State University, Ames, Iowa 50011, USA}
\newcommand{\jaea}{Advanced Science Research Center, Japan Atomic Energy Agency, 2-4 Shirakata Shirane, Tokai-mura, Naka-gun, Ibaraki-ken 319-1195, Japan}
\newcommand{\jeonbuk}{Jeonbuk National University, Jeonju, 54896, Korea}
\newcommand{\jyvaskyla}{Helsinki Institute of Physics and University of Jyv{\"a}skyl{\"a}, P.O.Box 35, FI-40014 Jyv{\"a}skyl{\"a}, Finland}
\newcommand{\kek}{KEK, High Energy Accelerator Research Organization, Tsukuba, Ibaraki 305-0801, Japan}
\newcommand{\korea}{Korea University, Seoul 02841, Korea}
\newcommand{\kurchatov}{National Research Center ``Kurchatov Institute", Moscow, 123098 Russia}
\newcommand{\kyoto}{Kyoto University, Kyoto 606-8502, Japan}
\newcommand{\lawllnl}{Lawrence Livermore National Laboratory, Livermore, California 94550, USA}
\newcommand{\losalamos}{Los Alamos National Laboratory, Los Alamos, New Mexico 87545, USA}
\newcommand{\lund}{Department of Physics, Lund University, Box 118, SE-221 00 Lund, Sweden}
\newcommand{\lyon}{IPNL, CNRS/IN2P3, Univ Lyon, Université Lyon 1, F-69622, Villeurbanne, France}
\newcommand{\maryland}{University of Maryland, College Park, Maryland 20742, USA}
\newcommand{\mass}{Department of Physics, University of Massachusetts, Amherst, Massachusetts 01003-9337, USA}
\newcommand{\michigan}{Department of Physics, University of Michigan, Ann Arbor, Michigan 48109-1040, USA}
\newcommand{\muhlenberg}{Muhlenberg College, Allentown, Pennsylvania 18104-5586, USA}
\newcommand{\nara}{Nara Women's University, Kita-uoya Nishi-machi Nara 630-8506, Japan}
\newcommand{\natmephi}{National Research Nuclear University, MEPhI, Moscow Engineering Physics Institute, Moscow, 115409, Russia}
\newcommand{\newmex}{University of New Mexico, Albuquerque, New Mexico 87131, USA}
\newcommand{\nmsu}{New Mexico State University, Las Cruces, New Mexico 88003, USA}
\newcommand{\northcg}{Physics and Astronomy Department, University of North Carolina at Greensboro, Greensboro, North Carolina 27412, USA}
\newcommand{\ohio}{Department of Physics and Astronomy, Ohio University, Athens, Ohio 45701, USA}
\newcommand{\ornl}{Oak Ridge National Laboratory, Oak Ridge, Tennessee 37831, USA}
\newcommand{\orsay}{IPN-Orsay, Univ.~Paris-Sud, CNRS/IN2P3, Universit\'e Paris-Saclay, BP1, F-91406, Orsay, France}
\newcommand{\peking}{Peking University, Beijing 100871, People's Republic of China}
\newcommand{\pnpi}{PNPI, Petersburg Nuclear Physics Institute, Gatchina, Leningrad region, 188300, Russia}
\newcommand{\pusan}{Pusan National University, Pusan 46241, Korea}
\newcommand{\riken}{RIKEN Nishina Center for Accelerator-Based Science, Wako, Saitama 351-0198, Japan}
\newcommand{\rikjrbrc}{RIKEN BNL Research Center, Brookhaven National Laboratory, Upton, New York 11973-5000, USA}
\newcommand{\rikkyo}{Physics Department, Rikkyo University, 3-34-1 Nishi-Ikebukuro, Toshima, Tokyo 171-8501, Japan}
\newcommand{\saispbstu}{Saint Petersburg State Polytechnic University, St.~Petersburg, 195251 Russia}
\newcommand{\seoulnat}{Department of Physics and Astronomy, Seoul National University, Seoul 151-742, Korea}
\newcommand{\stonybrkc}{Chemistry Department, Stony Brook University, SUNY, Stony Brook, New York 11794-3400, USA}
\newcommand{\stonycrkp}{Department of Physics and Astronomy, Stony Brook University, SUNY, Stony Brook, New York 11794-3800, USA}
\newcommand{\tenn}{University of Tennessee, Knoxville, Tennessee 37996, USA}
\newcommand{\texsu}{Texas Southern University, Houston, TX 77004, USA}
\newcommand{\titech}{Department of Physics, Tokyo Institute of Technology, Oh-okayama, Meguro, Tokyo 152-8551, Japan}
\newcommand{\tsukuba}{Tomonaga Center for the History of the Universe, University of Tsukuba, Tsukuba, Ibaraki 305, Japan}
\newcommand{\vandy}{Vanderbilt University, Nashville, Tennessee 37235, USA}
\newcommand{\weizmann}{Weizmann Institute, Rehovot 76100, Israel}
\newcommand{\wigner}{Institute for Particle and Nuclear Physics, Wigner Research Centre for Physics, Hungarian Academy of Sciences (Wigner RCP, RMKI) H-1525 Budapest 114, POBox 49, Budapest, Hungary}
\newcommand{\yonsei}{Yonsei University, IPAP, Seoul 120-749, Korea}
\newcommand{\zagreb}{Department of Physics, Faculty of Science, University of Zagreb, Bijeni\v{c}ka c.~32 HR-10002 Zagreb, Croatia}
\affiliation{\abilene}
\affiliation{\augie}
\affiliation{\banaras}
\affiliation{\barc}
\affiliation{\baruch}
\affiliation{\bnlcoll}
\affiliation{\bnlphys}
\affiliation{\caucr}
\affiliation{\charlesczech}
\affiliation{\ciae}
\affiliation{\cns}
\affiliation{\colorado}
\affiliation{\columbia}
\affiliation{\czechtech}
\affiliation{\debrecen}
\affiliation{\elte}
\affiliation{\eszterhazy}
\affiliation{\ewha}
\affiliation{\famu}
\affiliation{\fsu}
\affiliation{\gsu}
\affiliation{\hiroshima}
\affiliation{\howard}
\affiliation{\ihepprot}
\affiliation{\illuiuc}
\affiliation{\inrras}
\affiliation{\instpasczech}
\affiliation{\isu}
\affiliation{\jaea}
\affiliation{\jeonbuk}
\affiliation{\jyvaskyla}
\affiliation{\kek}
\affiliation{\korea}
\affiliation{\kurchatov}
\affiliation{\kyoto}
\affiliation{\lawllnl}
\affiliation{\losalamos}
\affiliation{\lund}
\affiliation{\lyon}
\affiliation{\maryland}
\affiliation{\mass}
\affiliation{\michigan}
\affiliation{\muhlenberg}
\affiliation{\nara}
\affiliation{\natmephi}
\affiliation{\newmex}
\affiliation{\nmsu}
\affiliation{\northcg}
\affiliation{\ohio}
\affiliation{\ornl}
\affiliation{\orsay}
\affiliation{\peking}
\affiliation{\pnpi}
\affiliation{\pusan}
\affiliation{\riken}
\affiliation{\rikjrbrc}
\affiliation{\rikkyo}
\affiliation{\saispbstu}
\affiliation{\seoulnat}
\affiliation{\stonybrkc}
\affiliation{\stonycrkp}
\affiliation{\tenn}
\affiliation{\texsu}
\affiliation{\titech}
\affiliation{\tsukuba}
\affiliation{\vandy}
\affiliation{\weizmann}
\affiliation{\wigner}
\affiliation{\yonsei}
\affiliation{\zagreb}
\author{U.A.~Acharya} \affiliation{\gsu} 
\author{A.~Adare} \affiliation{\colorado} 
\author{C.~Aidala} \affiliation{\michigan} 
\author{N.N.~Ajitanand} \altaffiliation{Deceased} \affiliation{\stonybrkc} 
\author{Y.~Akiba} \email[PHENIX Spokesperson: ]{akiba@rcf.rhic.bnl.gov} \affiliation{\riken} \affiliation{\rikjrbrc} 
\author{M.~Alfred} \affiliation{\howard} 
\author{V.~Andrieux} \affiliation{\michigan} 
\author{K.~Aoki} \affiliation{\kek} \affiliation{\riken} 
\author{N.~Apadula} \affiliation{\isu} \affiliation{\stonycrkp} 
\author{H.~Asano} \affiliation{\kyoto} \affiliation{\riken} 
% ...
\collaboration{PHENIX Collaboration}  \noaffiliation

\date{\today}

\begin{abstract}

There is strong evidence for the formation of small droplets of 
quark-gluon plasma in $p/d/^{3}$He+Au collisions at the Relativistic 
Heavy Ion Collider (RHIC) and in $p$+$p$/Pb collisions at the Large 
Hadron Collider.  In particular, the analysis of data at RHIC for 
different geometries obtained by varying the projectile size and shape 
has proven insightful.  In the present analysis, we find excellent 
agreement with the previously published PHENIX at RHIC results on 
elliptical and triangular flow with an independent analysis via the 
two-particle correlation method, which has quite different systematic 
uncertainties and an independent code base.  In addition, the results 
are extended to other detector combinations with different kinematic 
(pseudorapidity) coverage.  These results provide additional constraints 
on contributions from nonflow and longitudinal decorrelations.

\end{abstract}

\maketitle

\section{Introduction}

The Relativistic Heavy Ion Collider (RHIC) was built and the Large 
Hadron Collider (LHC) heavy-ion program initiated to study the formation 
of nucleus-sized droplets of quark-gluon plasma (QGP) in the laboratory. 
This focused scientific enterprise has been remarkably successful. The 
now standard model of heavy-ion collisions includes the formation of QGP 
that expands hydrodynamically before the phase transition to hadrons 
with confined quarks and 
gluons.  Refs.~\cite{Busza:2018rrf,Muller:2012zq,Muller:2006ee} provide 
useful reviews.  Over the past ten years, experiments have employed 
multiple techniques to assess whether such QGP droplets are also formed 
in smaller collisions of \pa and even \pp collisions; see 
Ref.~\cite{Nagle:2018nvi} for a recent review.

A specific proposal was to collide proton, deuteron, and helium-3 
projectiles on nuclear targets at RHIC, utilizing the unique 
capabilities of that facility, to discern whether ``flow-like'' patterns 
are indeed attributable to mini-QGP droplet 
formation~\cite{Nagle:2013lja}.  In the years 2014, 2015, and 2016, RHIC 
provided collisions of \heau, \pau, and \dau at \snn = 200 GeV, 
extending earlier results from 2003 and 2008 \dau running. The 
PHENIX Collaboration has published a suite of results on small systems 
including transverse momentum (\pt) spectra of identified particles 
(indicating ``baryon anomaly'' results in small collision 
systems)~\cite{Adare:2013esx}, pseudorapidity dependence of particle 
production and anisotropy coefficients~\cite{Adare:2018toe, 
Aidala:2017pup}, multiparticle cumulants~\cite{Aidala:2017ajz}, and 
anisotropy coefficients at midrapidity as a function of \pt for charged 
hadrons~\cite{Aidala:2016vgl,Adare:2015ctn,Adare:2014keg,Adare:2013piz} 
and for identified particles~\cite{Adare:2017rdq,Adare:2017wlc}.  The 
full set of elliptic and triangular azimuthal anisotropy coefficients 
($v_{2}$ and $v_{3}$, respectively) for all three collision geometries 
were published in Nature Physics~\cite{PHENIX:2018lia}.

The elliptical and triangular azimuthal anisotropy coefficients in all 
three collision geometries are quantitatively predicted by viscous 
hydrodynamic calculations published prior to the 
data~\cite{Nagle:2013lja,PhysRevC.95.014906}. After intense theoretical 
work~\cite{Mace:2018vwq,Mace:2018yvl} and intense scientific 
scrutiny~\cite{Nagle:2018ybc}, calculations with initial-state 
correlations in the color glass condensate framework are definitively 
ruled out as the dominant source of the observed correlations.  More 
recent calculations indicate that prehydrodynamization evolution, 
either in the weak~\cite{Schenke:2019pmk} or 
strong~\cite{Romatschke:2015gxa} coupling limit, may have a significant 
impact on the shortest lifetime systems---including the smallest systems 
or any size system at the lowest energies---particularly for triangular 
flow. Additionally, calculations within parton transport frameworks such 
as AMPT qualitatively reproduce the flow coefficient 
ordering~\cite{Koop:2015wea}. Finally, the initial geometry has 
contributions from both intrinsic geometry and from geometric 
fluctuations (originating from nucleonic and subnucleonic position-space 
fluctuations), and from the statistics of particle production. As an 
example, the initial spatial eccentricities $\varepsilon_{2}$ and 
$\varepsilon_{3}$ for central collisions (impact parameter 
$b<2~{\rm fm}$) in different frameworks are given in 
Table~\ref{tab:geometry}.  Additional negative binomial distribution 
fluctuations in particle production and sub-nucleonic structure tend to 
increase the eccentricities overall and reduce the differences, 
i.e.~reducing the relative contribution from intrinsic geometry; 
however, significant intrinsic contributions remain in almost all cases. 
Additional measurements and theoretical work are needed to gain insight 
on the relative contributions to the initial geometry and thus further 
constrain both the hydrodynamic and prehydrodynamization stages.

\begin{table}[hbt]
\caption{
Summary of various initial geometry calculations quantified by the 
average eccentricities $\varepsilon_{2,3}$ in central (impact parameter 
$b<2~{\rm fm}$) \pau, \dau, and \heau events.  The "Nucl. w/o NBD Fluc." 
column refers to Monte Carlo Glauber with nucleon position 
fluctuations~\cite{Nagle:2013lja}.  The "Nucl. w/ NBD Fluc." column 
refers to Monte Carlo Glauber with nucleon position fluctuations and 
Negative Binomial Distribution (NBD) fluctuations in particle 
production~\cite{Welsh:2016siu}.  The "Quarks w/ NBD Fluc." column 
refers to Monte Carlo Glauber with constituent quark position 
fluctuations and NBD fluctuations~\cite{Welsh:2016siu}.  The last two 
columns use the IP-Glasma framework including gluon field 
fluctuations~\cite{Schenke:2012wb} in publicly available code with 
nucleon and constituent quark position fluctuations.
}
\begin{ruledtabular}
\begin{tabular}{ccccccc}
 &           & Nucl.& Nucl. & Quarks & IP-G & IP-G \\
$\left< \varepsilon_{2,3} \right>$
 & Collision & w/o  & w/    & w/     & w/     & w/  \\
 & system    & NBD  &  NBD  &  NBD   &  Nucl. &  Quarks \\
 &           & Fluc. & Fluc. &  Fluc.&        &   \\ \hline
$\left< \varepsilon_{2} \right>$
 &  $p$$+$Au & 0.23 & 0.32 & 0.38 & 0.10 &  0.50 \\
 &  $d$$+$Au & 0.54 & 0.48 & 0.51 & 0.58 & 0.73 \\
 &  \heau    & 0.50 & 0.50 & 0.52 & 0.55 & 0.64 \\ 
\\
$\left< \varepsilon_{3} \right>$
 &  $p$$+$Au & 0.16 & 0.24 & 0.30 & 0.09 & 0.32 \\
 &  $d$$+$Au & 0.18 & 0.28 & 0.31 & 0.28 & 0.40 \\
 &     \heau & 0.28 & 0.32 &  0.35 &  0.34 & 0.46 \\
\end{tabular}
\end{ruledtabular}
\label{tab:geometry}
\end{table}

\begin{figure}[tbh]
\begin{minipage}{0.99\linewidth}
    \includegraphics[width=1.0\linewidth]{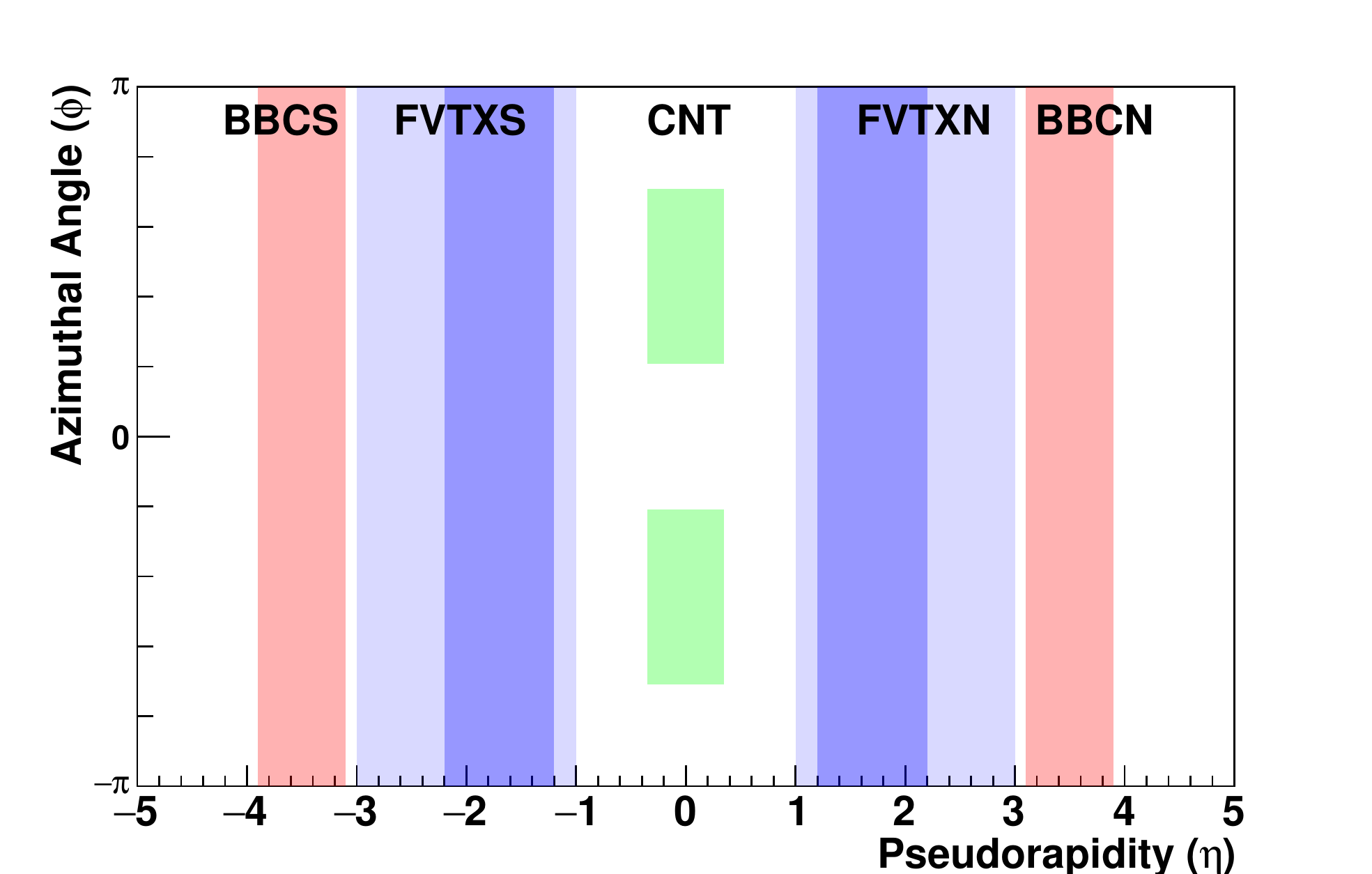}
\caption{PHENIX detector system layout.  The Beam-Beam Counters (BBCS 
and BBCN), the Forward Vertex Tracker (FVTXS and FVTXN), and the central 
spectrometer arms (CNT) are shown with their respective pseudorapidity 
coverage (horizontal) and azimuthal coverage (vertical).  For the FVTX, 
the lighter shaded region corresponds to the cluster acceptance while 
the darker shaded region corresponds to the reconstructed track 
acceptance.}
    \label{fig:layout}
\end{minipage}
\begin{minipage}{0.99\linewidth}
\vspace{-0.4cm}
    \includegraphics[width=1.0\linewidth]{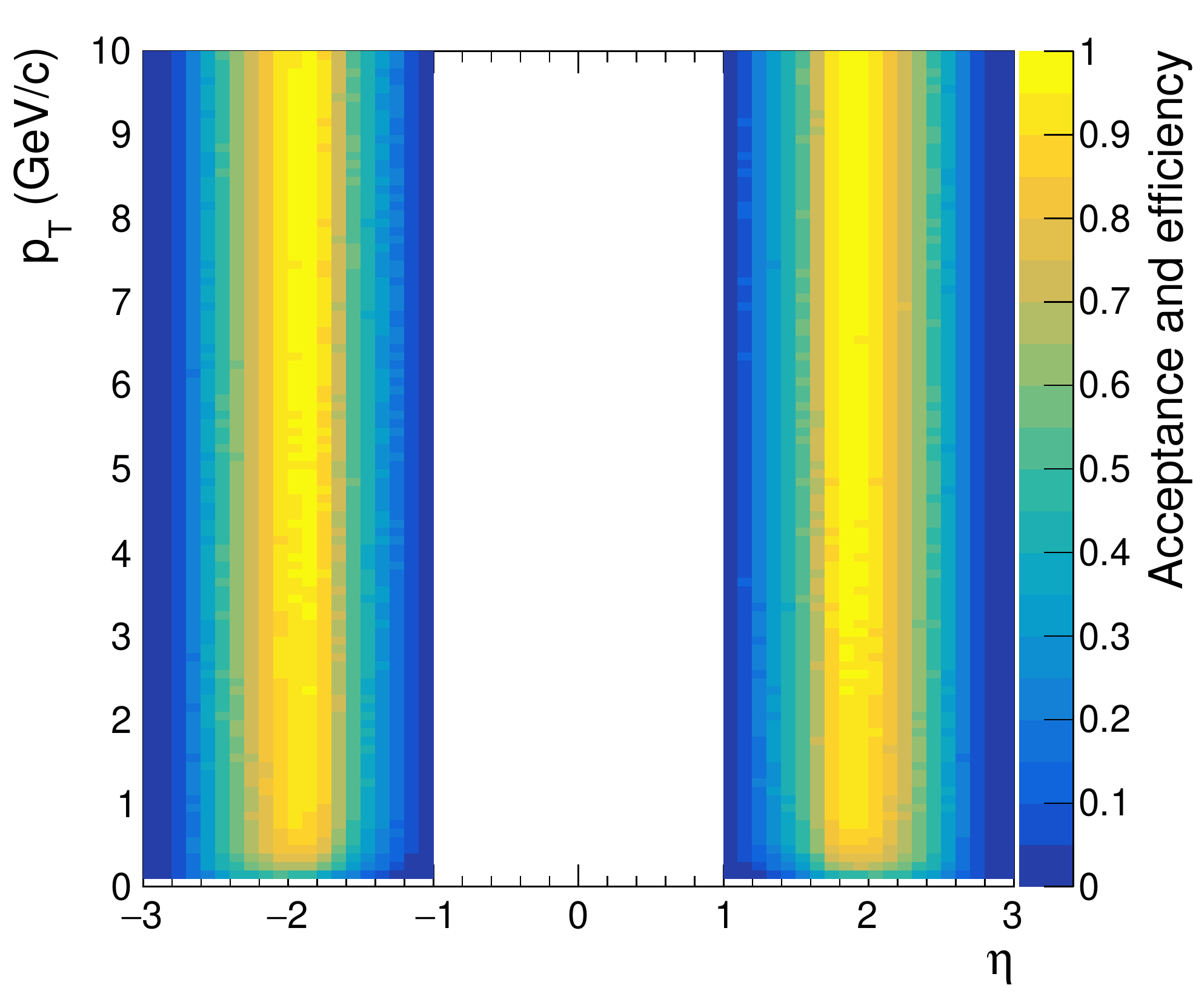}
\caption{PHENIX relative acceptance and efficiency for reconstructed 
tracks in the FVTX as a function of pseudorapidity and transverse 
momentum ($p_{T}$).  The pseudorapidity and $p_{T}$ acceptances depend 
on the collision $z$-vertex, and the $z$-vertex distribution is included 
in generating this map.  The tracking quality selection significantly 
reduces the efficiency for tracks with $p_{T} < 0.5~{\rm GeV}/c$, 
again pseudorapidity dependent.}
    \label{fig:fvtx_accept}
\end{minipage}
\end{figure}

Given the importance of these results, the PHENIX Collaboration has 
carried out a new analysis of the same data sets using combinations of 
three sets of detector combinations to extract two-particle 
correlations (2PC), called the \ttpc method, to check the published 
results~\cite{PHENIX:2018lia} and to provide additional information via 
correlations between particles from different kinematic regions. Because 
this makes use of three different two-particle correlations, it is 
called the \ttpc method. In addition, as the PHENIX experiment collected 
its final data in 2016, we provide an archival set of correlation 
function data for future examination. In this paper, we do not compare 
the experimental results with the latest theoretical calculations and 
rather focus solely on the measurements and their quantified 
uncertainties.

\section{Analysis Method}

The following subsections detail the PHENIX detector and the 
correlation analysis.

\subsection{Detector description}

The PHENIX detector is composed of multiple spectrometers and detector 
subsystems~\cite{Adcox:2003zm,Akikawa:2003zs}.  The detectors used in 
this analysis are highlighted in Fig.~\ref{fig:layout} and detailed 
here.  The central arm spectrometers (CNT) measure charged hadrons with 
pseudorapidity $|\eta|<0.35$.  There are two CNT spectrometers, referred 
to as ``east'' and ``west'', each subtending $\Delta\phi = \pi/2$. The 
beam-beam counters (BBC)~\cite{ALLEN2003549} comprise two sets of 64 
quartz \v{C}erenkov radiators with photomultiplier readout, each set 
covering $3.1 < |\eta| < 3.9$---the BBC covering $-3.9<\eta<-3.1$ is 
referred to as the ``south'' side (BBCS), and likewise the BBC covering 
$3.1<\eta<3.9$ is called the ``north'' side (BBCN). No individual 
particle information is available and the light output for each counter 
is normalized to the expected single charged particle response. We note 
that approximately half of the particles hitting the BBC are scattered 
from the beam pipe and the poles of the axial field magnet. The forward 
silicon vertex detector (FVTX)~\cite{Aidala:2013vna} comprises silicon 
strips oriented in the azimuthal direction and covers both forward and 
backward rapidity $1.0 < |\eta| < 3.0$.  The FVTX can be used to count 
hits via clusters or via reconstructed tracks in the four-layers on each 
side. The acceptance for FVTX tracks is significantly more constrained 
than the acceptance for clusters, and has a strong dependence on the 
$z$-vertex of the collision (the direction along the beam line).  Due to 
the orientation of the strips, there is no momentum information 
available with the FVTX tracks.  The FVTX acceptance for tracks is shown 
in Fig.~\ref{fig:fvtx_accept} and is dominated by tracks with 
$1.2<|\eta|<2.2$ and $p_{T}>0.5~{\rm GeV}/c$.

The BBC is used for triggering on minimum bias (MB) \pau, \dau, and \heau 
collisions by requiring a fast reconstructed $z$-vertex within 
$|z|<10$~cm and at least one hit on each side of the collision point.  
Additionally, a high-multiplicity trigger was employed to enhance the 
0\%--5\% highest BBC multiplicity events by more than an order of 
magnitude. The BBC information in the Au-going direction is also used 
offline to select events in the 0\%--5\% centrality category.  Full 
details are available in 
Refs.~\cite{PHENIX:2018lia,Adare:2017rdq,Aidala:2016vgl,Adare:2015ctn}.

\subsection{Event plane method}

Previous PHENIX publications, including Ref.~\cite{PHENIX:2018lia}, 
utilized the event plane method~\cite{Ollitrault:2009ie} for measuring 
azimuthal anisotropies. The second- and third-harmonic event planes are 
determined in the BBC in the Au-going direction (referred to as the BBC 
``south'' or BBCS) and in the Au-going FVTX (referred to as the FVTX 
``south'' or FVTXS). The standard Q-vector recentering and event plane 
flattening techniques~\cite{Poskanzer:1998yz} are applied. Because the 
collision system is asymmetric, one cannot determine the event plane 
resolutions by comparing forward and backward detectors alone. Thus, the 
event plane resolutions are determined utilizing the three-detector 
combination BBCS-FVTXS-CNT.

It was recently pointed out that the third-harmonic event plane 
resolutions for the BBCS and FVTXS published in 
Ref.~\cite{PHENIX:2018lia} do not follow the expected simple scaling of 
$R(\psi_n) \propto v_n\sqrt{N_{\rm hit}}$, where $N_{\rm hit}$ is the 
number of particles striking the event plane detector and $v_{n}$ is the 
azimuthal anisotropy of those particles. We have carefully investigated 
this observation by running a full simulation of the event plane 
procedure, including the fact that the beam has nonzero angle and offset 
with respect to the detector coordinate system.  The beam angles and 
offsets for the different running periods are given in 
Table~\ref{tab:angles}.  An additional issue is that the PHENIX central 
carriage, which was moved between operation periods, has modest position 
offsets of order 1--2 mm relative to nominal. We find that the event 
plane flattening procedure in the rotated frame creates a distortion on 
the triangular anisotropy due to the elliptic anisotropy. The simulation 
qualitatively reproduced the event plane ``bias'' seen in real data; the 
effect largely cancels in the final $v_3$, because the bias is opposite 
between the BBCS and FVTXS. The effect is dependent on the size of the 
real signal $v_3$, the beam angle, beam offset, event multiplicity, and 
Q-vector recentering applied, and is much larger in \pau and \dau, where 
the smaller $v_3$ induces higher sensitivity to these effects.

\begin{table}[hbt]
\caption{System beam angles and offsets.}
\begin{ruledtabular}
\begin{tabular}{ccccc}
Year & System & $x_{{\rm offset}}$ & $y_{{\rm offset}}$ & $x$-$z$ angle  \\
     &        & (mm)                & (mm)                & (mrad) \\ \hline
2014 & \heau  & 3.9                 & 0.02                & 1.8  \\
2015 & \pau   & 2.1                 & 0.5                 & 3.6  \\
2016 & \dau   & 3.0                 & 0.2                 & 1.0  \\
\end{tabular}
\end{ruledtabular}
\label{tab:angles}
\end{table}

Because the experimental results for $v_3$ in \pau and \dau, where the 
distortion is largest, are important, we have carried out an independent 
analysis to examine the validity of the previous results. In the Monte 
Carlo simulation, two-particle correlation functions were successfully 
obtained when using an event-mixing acceptance correction in very fine 
bins in collision $z$-vertex.  Thus, we have carried out a new analysis 
of all three collision systems using three sets of two-particle 
correlations (\ttpc).  In the limit of low event plane resolution, which 
is the case for all three systems, the event plane physics result and 
the \ttpc physics result should agree~\cite{Ollitrault:2009ie}---this is 
because they are both estimators of $\left<v_2\right>^{1/2}$ in this 
case, which means the sensitivity to both fluctuations and nonflow is 
the same.

We highlight that the analysis is independent of the published event 
plane results in the following ways: (1) a completely different code 
base is used; (2) the FVTX clusters are used in the event plane result 
but only FVTX tracks with good quality are used in the \ttpc analysis; 
(3) additional systematic uncertainty checks are carried out as detailed 
below.  Note that a subset of these \ttpc checks were carried out in 
the \dau published analyses detailed in 
Refs.~\cite{PHENIX:2018lia,Aidala:2017pup}. In this paper, we also 
extend the kinematics from the original analysis to utilize different 
combinations of detectors in the \ttpc method.

\subsection{\ttpc method}

Here we detail the methodology used for the \ttpc method. The 2PC 
technique utilized here follows the standard 
methodology~\cite{Ollitrault:2009ie}; the difference only coming in 
requiring three such 2PC because the collision systems are asymmetric. We 
measure the $\Delta\phi$ distribution of three different sets of pairs. 
In each pair, one particle is required to be in one subevent, and the 
other is required to be in another subevent. The manner in which the 
three different pairs of subevents are used is qualitatively very 
similar to the three subevent method for determining the event plane 
resolution. In the limit of small event plane resolution, the techniques 
should yield the same results as they are sensitive to flow, flow 
fluctuations, and nonflow in the same manner~\cite{Ollitrault:2009ie}.

The correlation function $C(\Delta\phi)$ is defined by
\begin{equation}
C(\Delta\phi) = \frac{S(\Delta\phi)}{M(\Delta\phi)}\frac{\int_0^{2\pi}M(\Delta\phi)}{\int_0^{2\pi}S(\Delta\phi)},
\end{equation}
where $\Delta\phi$ is the difference in the azimuthal angles between the 
two particles in the pair; $S(\Delta\phi)$ is the signal distribution, 
which is constructed from pairs in which both particles are taken from 
the same event; and $M(\Delta\phi)$, which is the mixed event 
distribution, which is constructed from pairs of particles in which each 
particle is required to be from a different event. It is essential that 
particles from mixed events come from the same \textit{event category}, 
which includes centrality class and collision $z$-vertex class (i.e.~the 
collision $z$-vertices of both particles must be in the same collision 
$z$-vertex bin, typically 1 cm or 2 cm in width).

Once the correlation function is obtained, it can be decomposed via a 
Fourier series with coefficients $c_n$:
\begin{align}
C(\Delta\phi) = 1 + \sum_{n=1} c_n\cos\Delta\phi,
\end{align}
where $n$ is the harmonic number.
Letting the superscripts denote subevents $A$, $B$, and $C$,
the $c_n$ coefficients mathematically represent
\begin{align}
c_n^{AB} &= \mean{\cos(n(\phi_A-\phi_B))} = \mean{v_n^Av_n^B}, \\
c_n^{AC} &= \mean{\cos(n(\phi_A-\phi_C))} = \mean{v_n^Av_n^C}, \\
c_n^{BC} &= \mean{\cos(n(\phi_B-\phi_C))} = \mean{v_n^Bv_n^C}.
\end{align}
Finally, the $v_n$ in a single subevent can be determined as
\begin{equation}
v_n^C(p_T) = \sqrt{ \frac{c_n^{AC}(p_T)c_n^{BC}(p_T)}{c_n^{AB}}  }.
\end{equation}
Note that it is also possible to determine the $v_n$ in a different way, 
using only one correlation in the numerator and all three in the 
denominator:
\begin{align}
v_n^C(p_T) &= \frac{c_n^{AC}(p_T)}{\sqrt{c_n^{AB}c_n^{AC}/c_n^{BC}}}, \\
v_n^C(p_T) &= \frac{c_n^{BC}(p_T)}{\sqrt{c_n^{AB}c_n^{BC}/c_n^{AC}}};
\end{align}
where all of the correlations in the denominator are $p_T$-integrated. 
For the detectors without momentum information (BBCS, FVTXS, FVTXN), 
this simply means all tracks or hits. For the detectors with momentum 
information (CNT), this means all tracks in the momentum range 
considered ($0.2 <p_T< 3.0~{\rm GeV}/c$). Because this method is 
sometimes used by the LHC experiments, we will informally refer to it as 
the LHC-style $v_n$, in contrast to the PHENIX-style discussed 
previously.  Note that the PHENIX-style $v_n$ is the geometric mean of 
the two possible LHC-style $v_n$. For that reason, the PHENIX-style 
presents certain advantages, particularly for reduced systematic 
uncertainties.

\subsection{Systematic uncertainties}

The systematic uncertainties on the extracted $v_{2}$ and $v_{3}$ 
coefficients have multiple contributions.  In previous analyses 
utilizing CNT 
tracks~\cite{PHENIX:2018lia,Adare:2017rdq,Aidala:2016vgl,Adare:2015ctn,Adare:2014keg}, 
contributions from variations in track quality criteria, run-to-run 
variations, etc.~are quite modest and subdominant.  The two dominant 
sources of systematic uncertainty result from comparing results with 
different collision $z$-vertex ranges and from comparing the two 
individual arms of the CNT.  The uncertainty associated with the 
collision $z$-vertex is assessed by comparing the nominal result with 
$|z| < 10~{\rm cm}$ to cases with $+4.0~{\rm cm} < z <+10.0~{\rm cm}$ 
and $-10.0~{\rm cm} < z < -4.0~{\rm cm}$, because the changes in the 
FVTXS acceptance are significant over this range.  We also consider the 
variation of only using ``east'' arm CNT tracks and only using ``west'' 
arm CNT tracks.  The differences found in the systematic variations are 
taken to be the maximal possible deviations, representing asymmetric 
distributions about the central value.  These differences are divided by 
$\sqrt{3}$ to give one standard deviation uncertainties, and then those 
individual uncertainties are added in quadrature.

For the results utilizing the FVTXS and FVTXN detectors, we have 
repeated the analyses using only one half, i.e. $1.2 < |\eta| < 1.7$, or 
the other half, i.e. $1.7 < |\eta| < 2.2$.  These give similar results; 
however, we do not include the differences in the systematic 
uncertainties as the results may have differing contributions from 
nonflow and longitudinal decorrelations.  These results are presented 
in the Appendix.

\section{Results}
\label{sec:results}

\begin{figure*}[hbt]
\begin{minipage}{0.965\linewidth}
\includegraphics[width=0.32\linewidth]{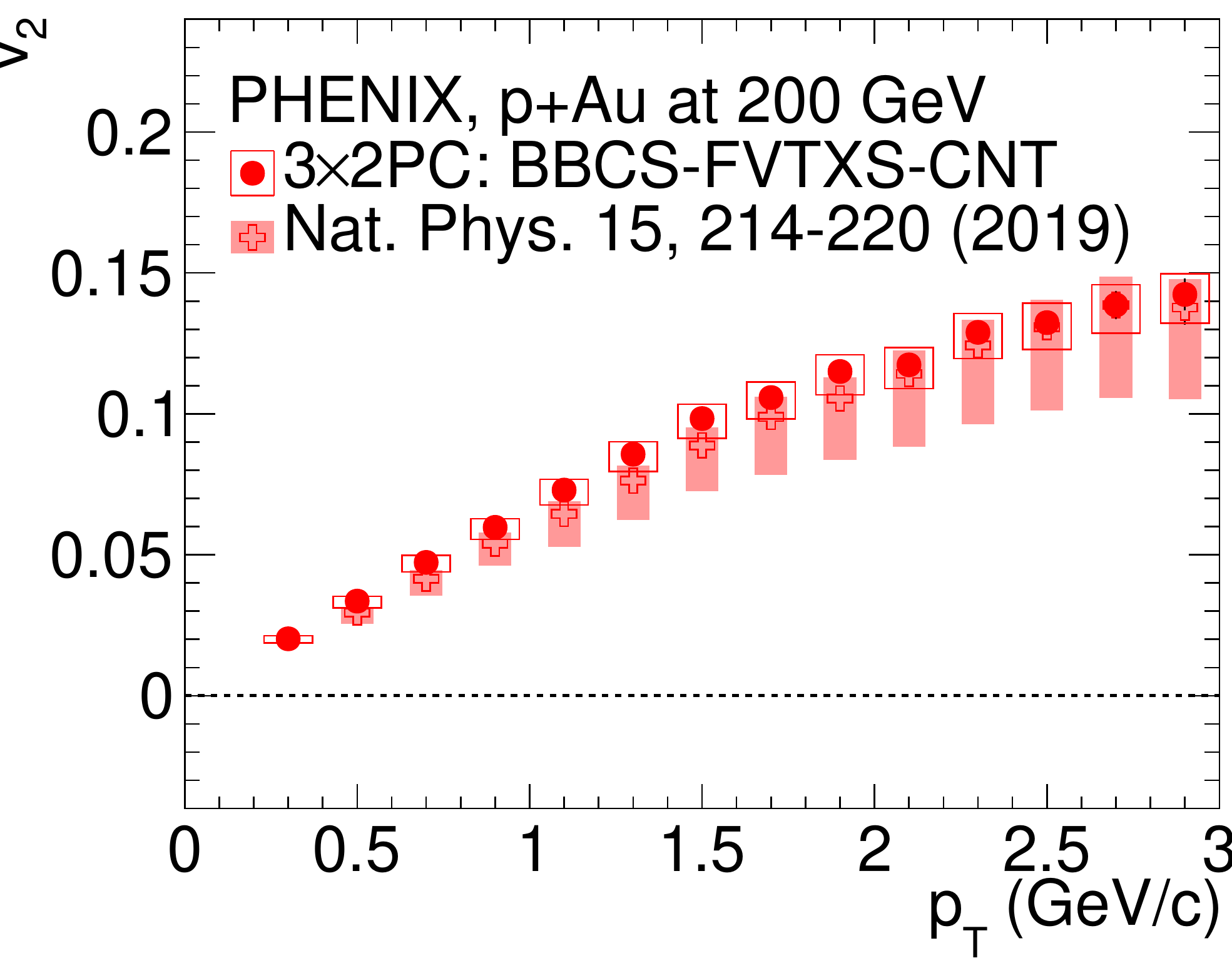}
\includegraphics[width=0.32\linewidth]{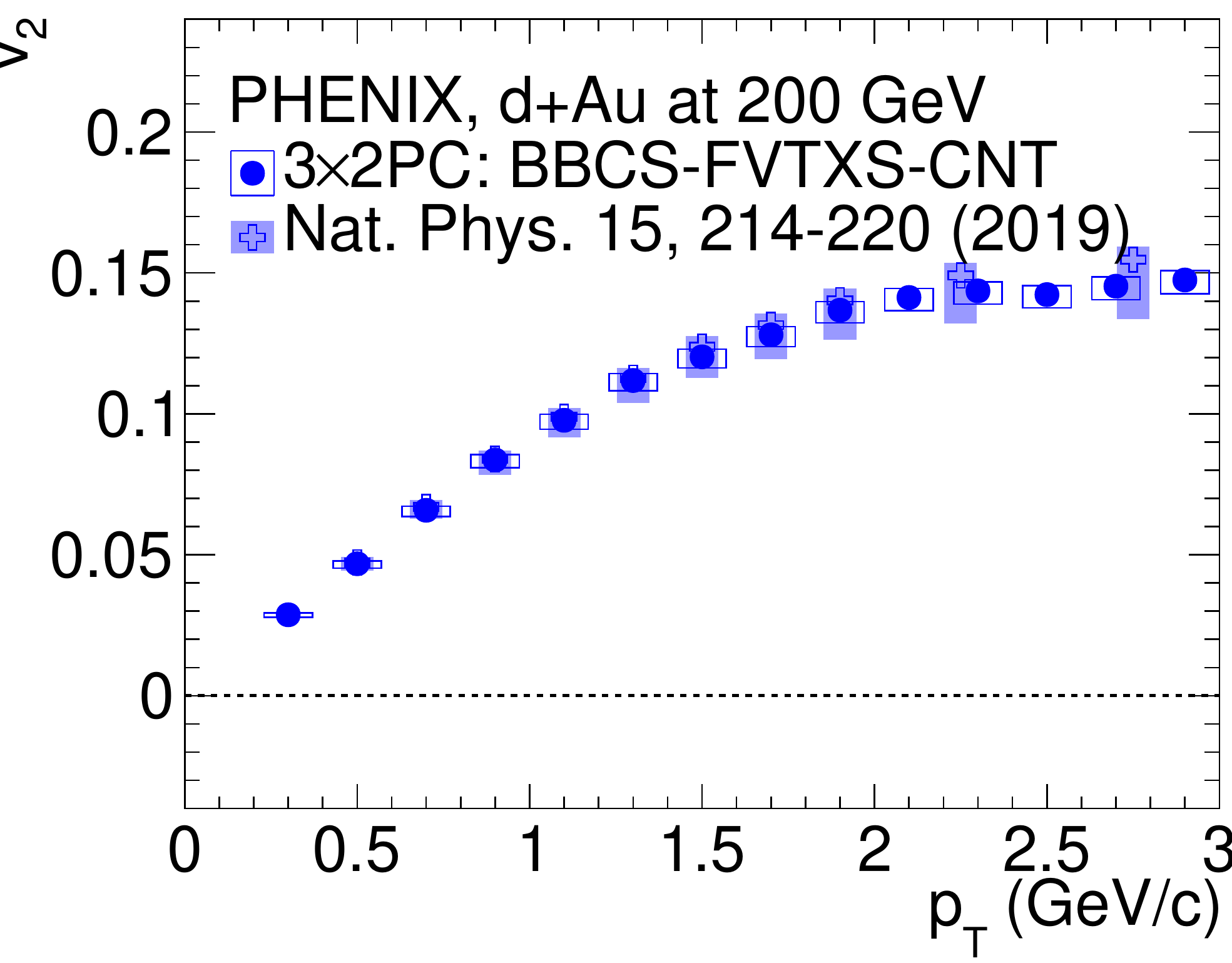}
\includegraphics[width=0.32\linewidth]{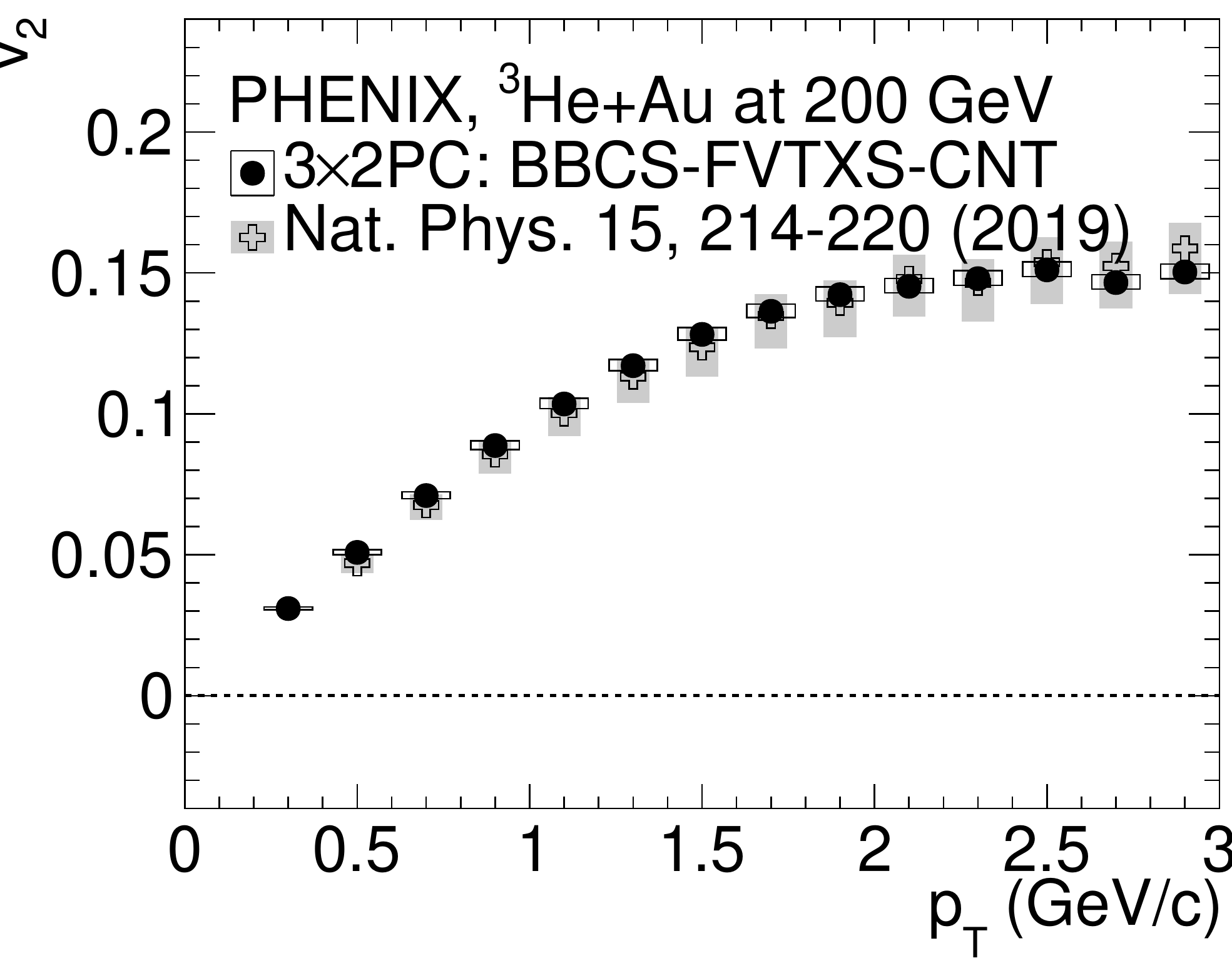}
\vspace{-0.4cm}
\caption{
The extracted $v_2$ coefficient as a function of \pt in 0\%--5\% central 
$p$$+$Au, $d$$+$Au, and $^3$He+Au collisions from 
the \ttpc method using the BBCS-FVTXS-CNT detector combination are shown 
as solid circles.  Also shown as open crosses are the published Nature 
Physics~\cite{PHENIX:2018lia} results that were obtained via the event 
plane method and the same combination of detectors.
}
\label{fig:res_ppg216_v2_bf}
\end{minipage}
\begin{minipage}{0.965\linewidth}
\includegraphics[width=0.32\linewidth]{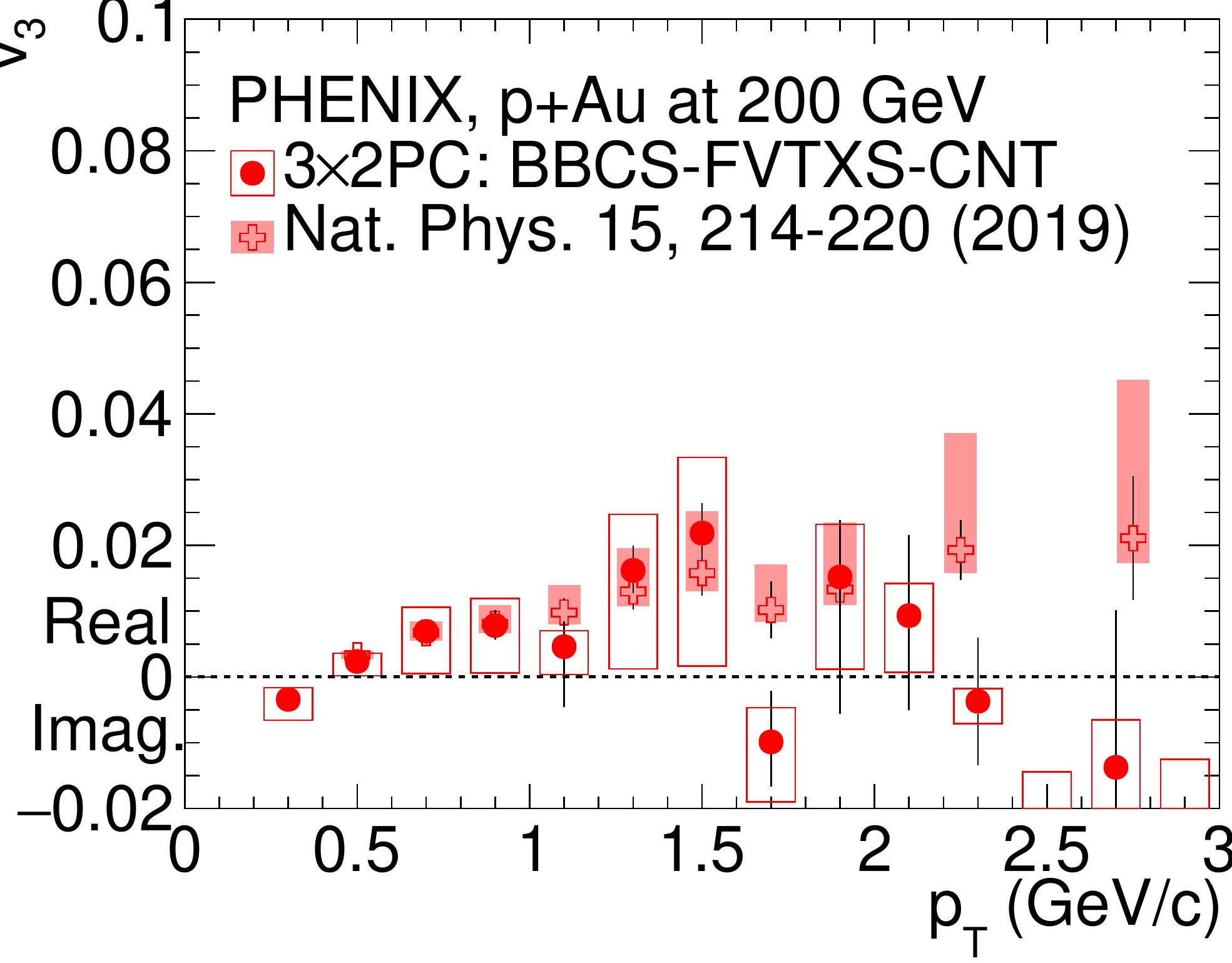}
\includegraphics[width=0.32\linewidth]{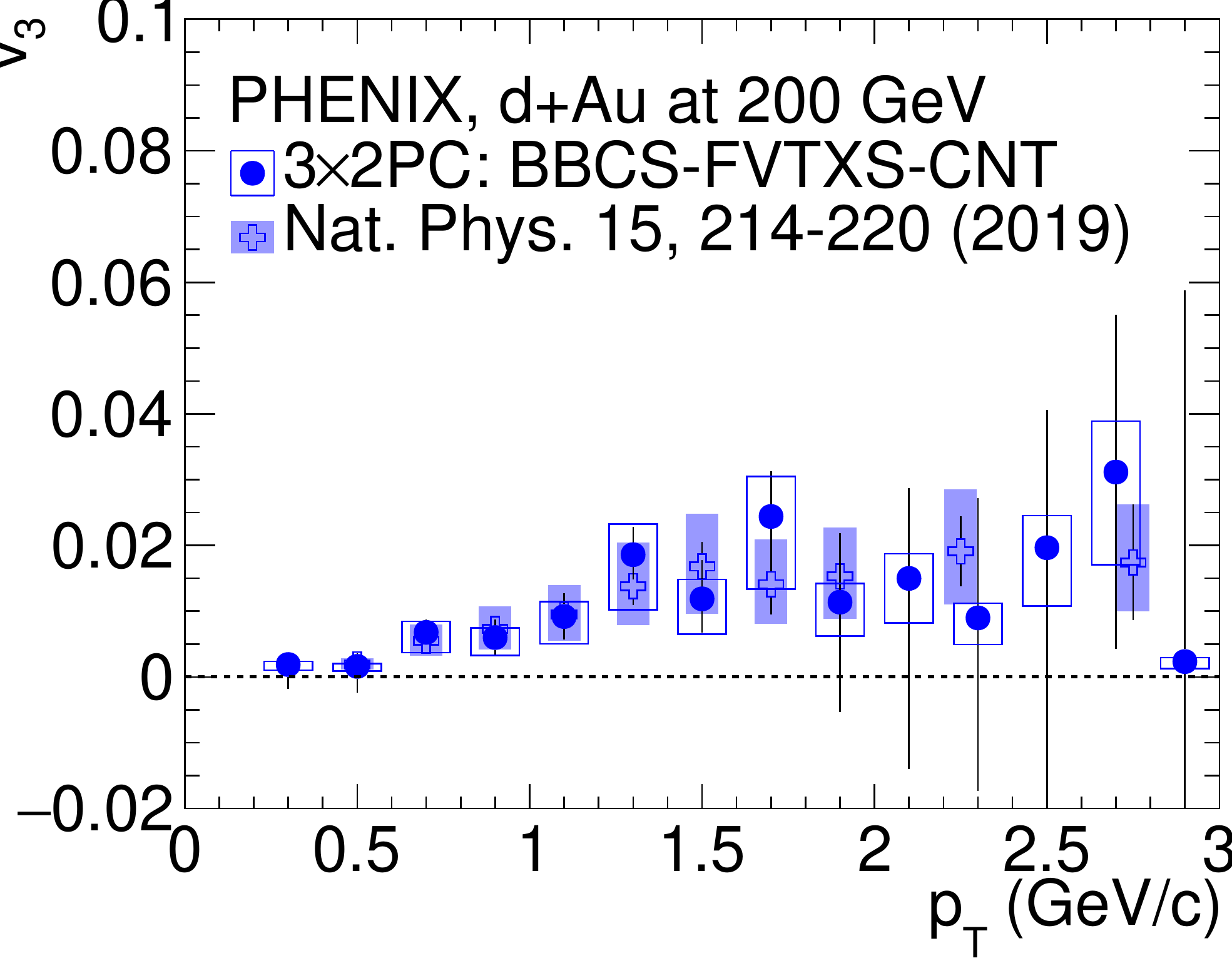}
\includegraphics[width=0.32\linewidth]{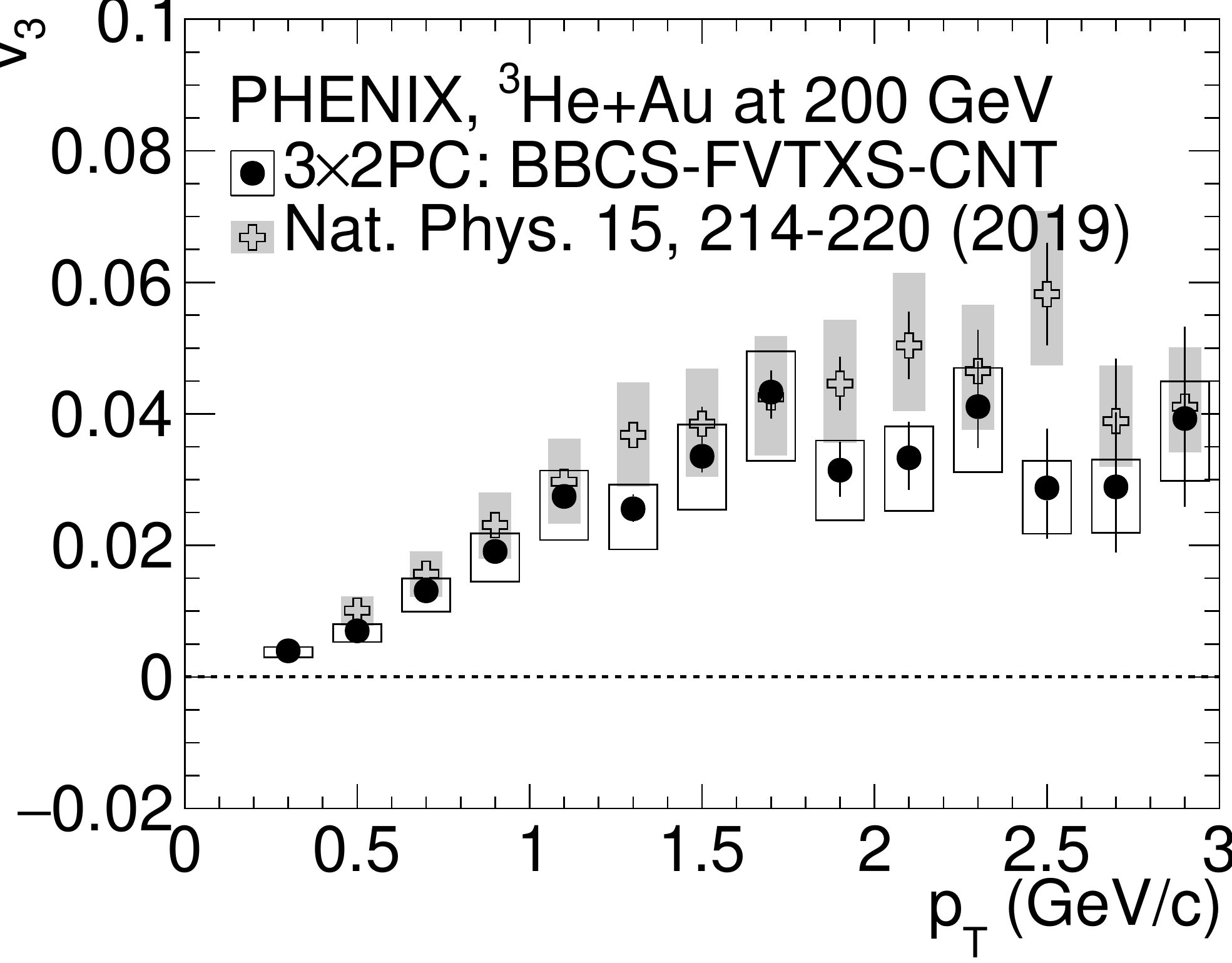}
\vspace{-0.4cm}
\caption{
The extracted $v_3$ coefficient as a function of \pt in 0\%--5\% central 
$p$$+$Au, $d$$+$Au, and $^3$He+Au collisions from 
the \ttpc method using the BBCS-FVTXS-CNT detector combination are shown 
as solid circles.  Also shown as open crosses are the published Nature 
Physics~\cite{PHENIX:2018lia} results that were obtained via the event 
plane method and the same combination of detectors.
}
\label{fig:res_ppg216_v3_bf}
\end{minipage}
\begin{minipage}{0.965\linewidth}
\includegraphics[width=0.32\linewidth]{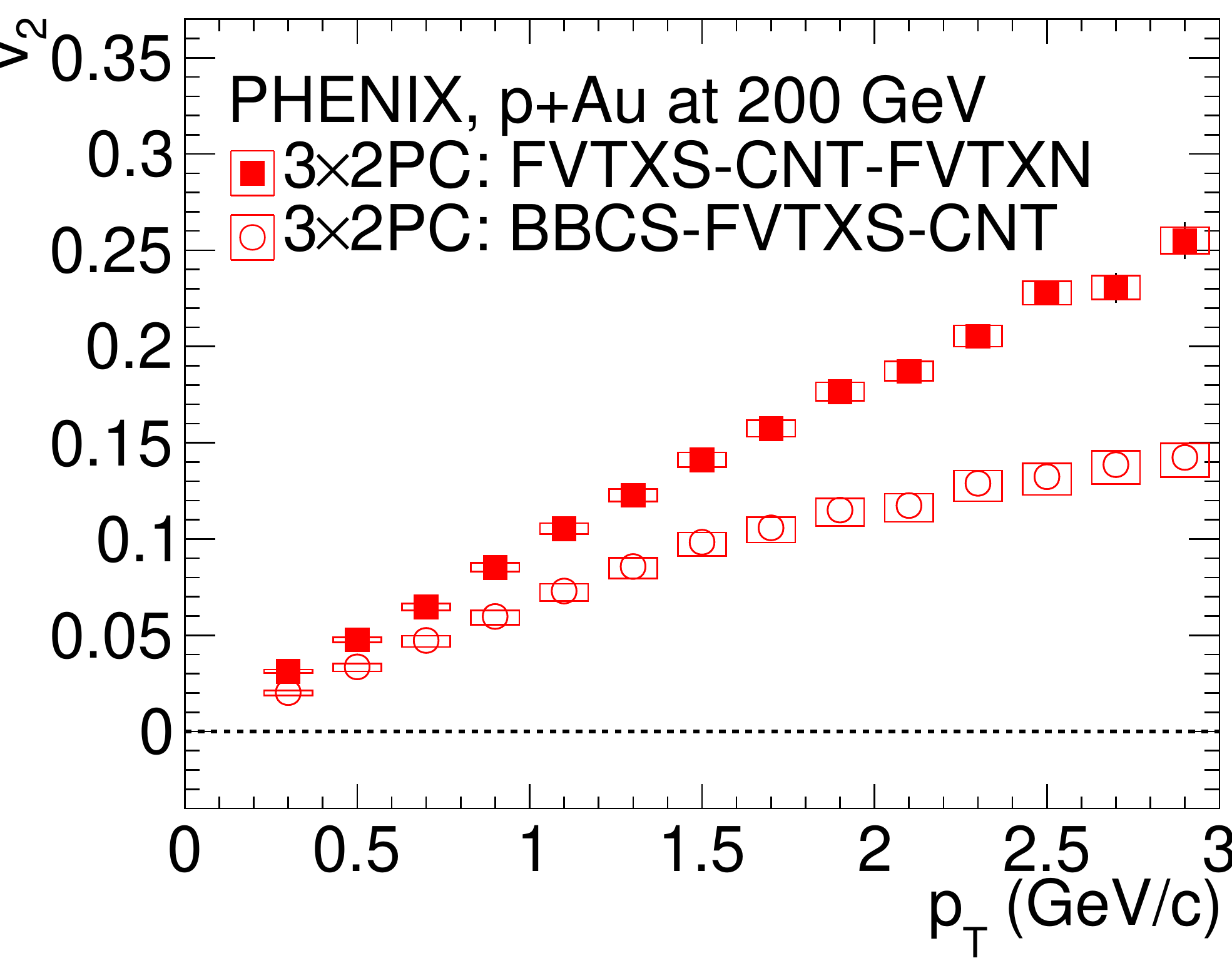}
\includegraphics[width=0.32\linewidth]{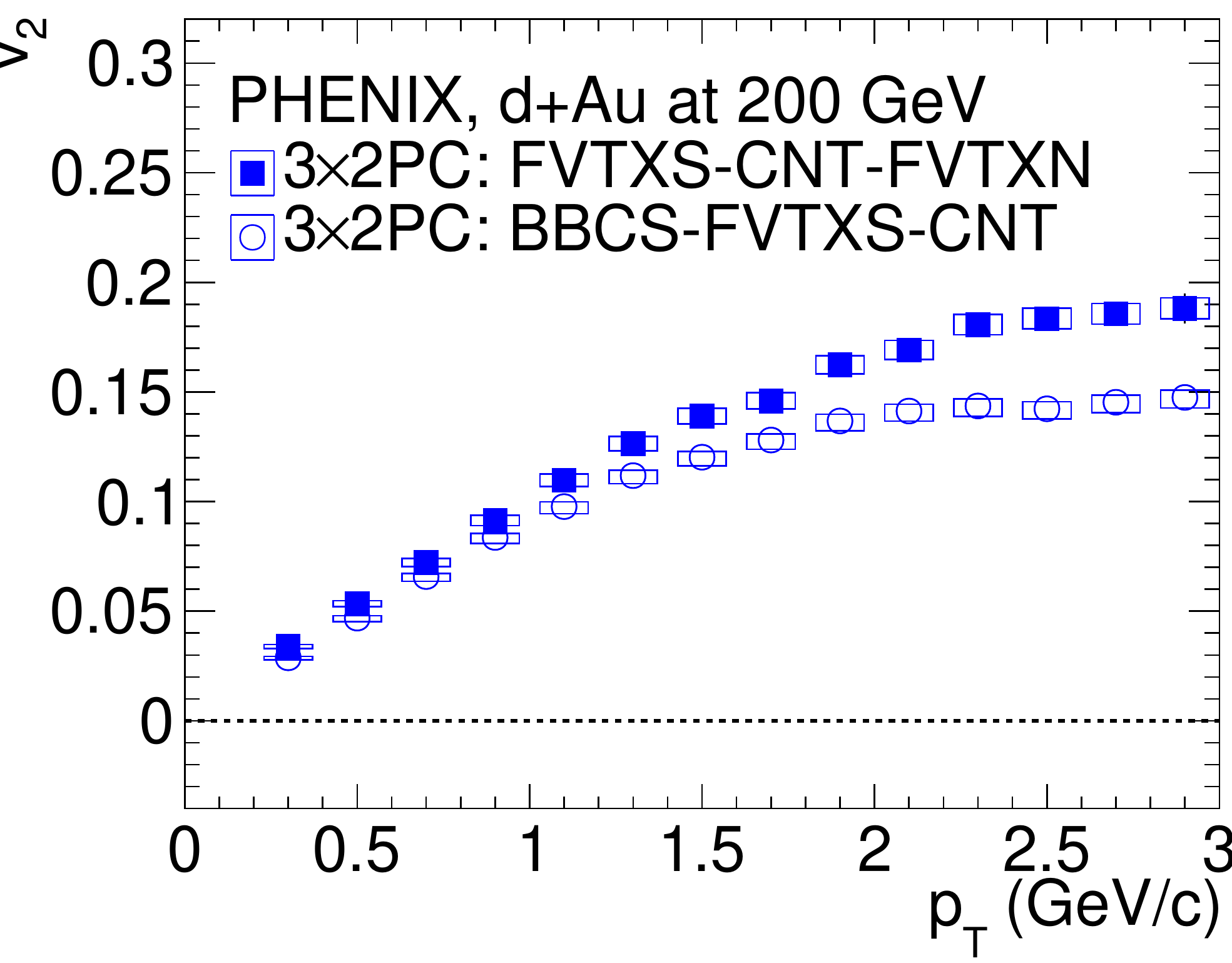}
\includegraphics[width=0.32\linewidth]{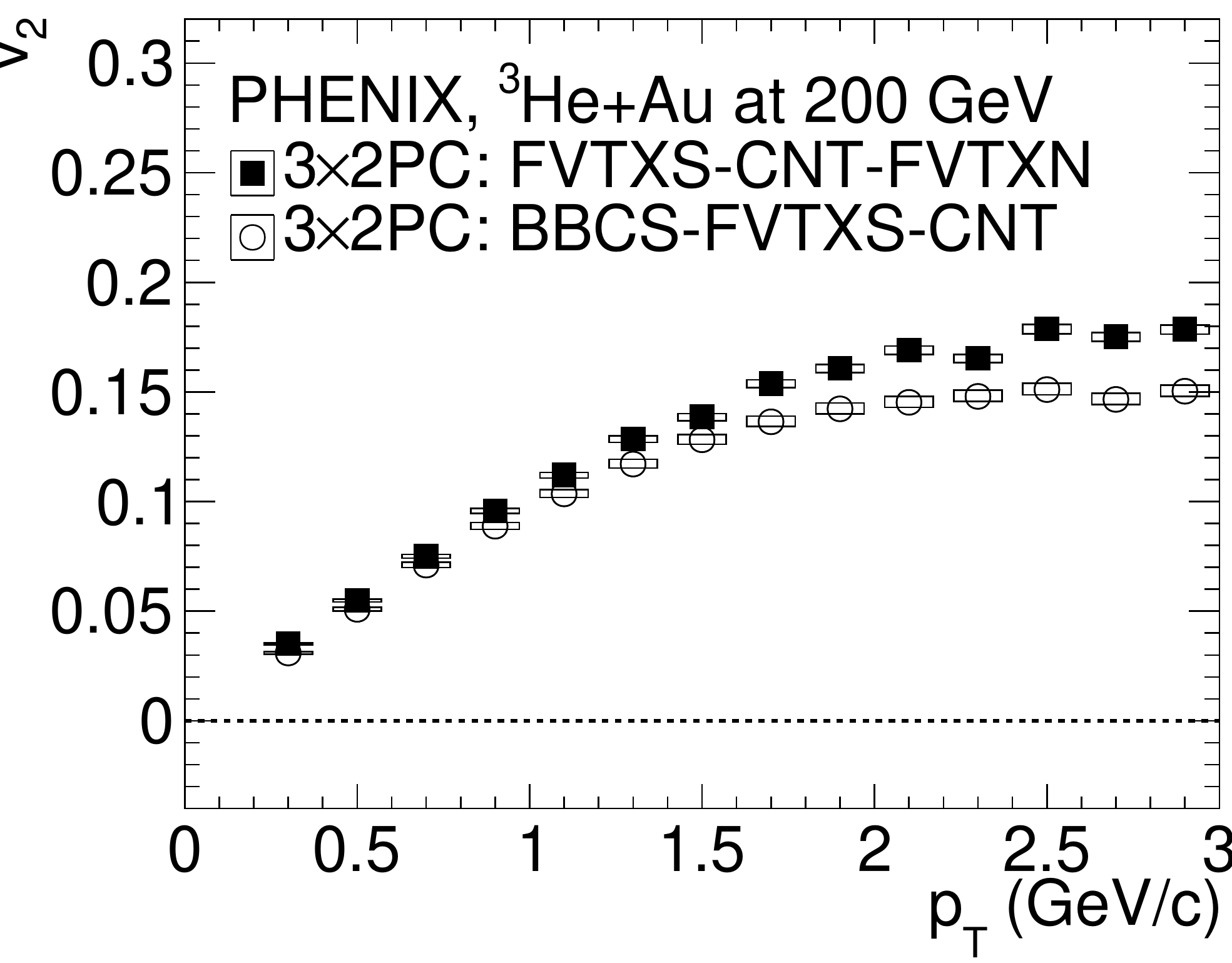}
\vspace{-0.4cm}
\caption{
The extracted $v_2$ coefficient as a function of \pt in 0\%--5\% central 
$p$$+$Au, $d$$+$Au, and $^3$He+Au collisions from 
the \ttpc method using the FVTXS-CNT-FVTXN detector combination are 
shown as solid squares.  For comparison we also show the previously 
plotted results from the BBCS-FVTXS-CNT combination as open circles.
}
\label{fig:res_ppg216_v2_ff}
\end{minipage}
\end{figure*}

The main physics results from the \ttpc analysis are the extracted 
$v_{2,3}$ coefficients as a function of charged hadron \pt at 
midrapidity $|\eta|<0.35$.  However, these values may depend on the 
other two detectors used in combination with the CNT tracks.  A set of 
example two-particle correlations and a complete set of extracted 
Fourier coefficients ($c_1$, $c_2$, $c_3$, $c_4$) and their statistical 
uncertainties are given in supplemental material~\cite{supp_mat}.

We highlight that care should be employed when comparing $c_n$ 
coefficients directly as the \pt acceptance of the BBCS and FVTXS 
differ, as well as their relative particle (direct and scattered) 
contributions.  Thus, even though in principle one can extract $v_{2}$ 
and $v_{3}$ values in the BBCS and FVTXS, by the same procedure as in 
the CNT, they do not have a straightforward physics interpretation.  In 
the case of reconstructed, high quality FVTXS tracks, a full acceptance 
and efficiency correction as a function of collision $z$-vertex is 
possible, as done for example in Ref.~\cite{Adare:2018toe}.  However, a 
similar procedure has not been done for the BBCS, where approximately 
half the hits are from scattered particles.

\subsection{$v_n$ vs $p_T$ results}

Figure~\ref{fig:res_ppg216_v2_bf} shows the elliptic $v_{2}$ 
coefficients as a function of \pt from the \ttpc method utilizing the 
three-detector combination BBCS-FVTXS-CNT.  The results for the most 
central 0\%--5\% events are shown for all three collision geometries with 
statistical uncertainties as vertical lines and systematic uncertainties 
as open boxes.  The systematic uncertainties have a high degree of 
point-to-point correlation. Also shown are the previously published 
$v_{2}$ coefficients~\cite{PHENIX:2018lia} utilizing the event plane 
method.  We highlight that the earlier publication includes an 
asymmetric systematic uncertainty estimate for nonflow based on a 
simple multiplicity scaling of coefficients from \pp collisions at \sqs 
= 200 GeV; here we do not include this uncertainty, as we focus on what 
is directly measured from the correlation functions with all physics 
contributions included.  The analysis presented here is in excellent 
agreement with the previously published results.

Figure~\ref{fig:res_ppg216_v3_bf} shows the third harmonic coefficient 
$v_3$ as a function of \pt.  It is otherwise identical to the previous 
figure, showing a comparison between the present \ttpc analysis and the 
previously published event plane analysis~\cite{PHENIX:2018lia} for the 
most central 0\%--5\% events for all three collision systems and, as 
before, the vertical lines represent the statistical uncertainties and 
the boxes indicate the point-to-point correlated systematic 
uncertainties. There is good agreement within uncertainties between the 
two analyses, with the \heau $v_{3}$ values from the \ttpc method 
slightly lower than the event plane results, though well within 
systematic uncertainties, which are largely independent between the two 
methods.


We highlight that the correlation coefficients $c_2$ and $c_3$ from all 
collision geometries follow the approximate expected scaling based on 
each detector's multiplicity and $v_n$ using inputs from 
Ref.~\cite{Adare:2018toe}.  Thus the puzzle involving the nonscaling of 
the event plane values mentioned above is resolved.

\subsection{Additional kinematic ranges}

The above results are presented as $v_{2}$ and $v_{3}$ at midrapidity 
$|\eta|<0.35$, but they can depend on the other two detectors used in 
the analysis, namely the BBCS and FVTXS.  Nonflow contributions, 
longitudinal decorrelations, and potentially other effects can make the 
extraction dependent on the kinematic coverage of the other detectors -- 
see for example Ref.~\cite{Lim:2019cys}.  The original motivation for 
utilizing the BBCS and FVTXS is based on their higher multiplicity and 
significant pseudorapidity gap from the CNT tracks.  Note that the gap 
should be thought of not in simple terms of the extreme $|\Delta \eta| > 
X$ value, but rather the distribution of possible $|\Delta \eta|$ 
values. Thus, the average $\left< |\Delta \eta| \right> \approx 3.5, 
2.0, 1.8$ for the BBCS-CNT, FVTXS-CNT, BBCS-FVTXS detector combinations.

We have also analyzed the detector combination FVTXS-CNT-FVTXN for the 
\ttpc.  Note that now the average values are $\left< |\Delta \eta| 
\right> \approx 2.0, 2.0, 3.4$ for the FVTXS-CNT, CNT-FVTXN, and 
FVTXS-FVTXN detector combinations.  However, based on 
measurements as a function of pseudorapidity in 
Ref.~\cite{Adare:2018toe}, the $v_{2}$ values are less than half the 
magnitude in the FVTXN compared to FVTXS and the multiplicity of tracks 
is also less than half.  Thus, nonflow contributions relative to flow 
contributions are expected to be substantially larger in the FVTXN.  
Again, additional example correlation functions and the full set of 
extracted $c_{n}$ coefficients are given in the Appendix.

Tables~\ref{tab:detaci} and \ref{tab:detacc} list the pseudorapidity 
acceptances of the different detectors and two-detector combinations.  
In Fig.~\ref{fig:res_ppg216_v2_bf} the pseudorapidity acceptance of the 
BBCS-FVTXS-CNT combination is listed as $-3.9<\eta<-3.1, -3.0<\eta<-1.0, 
|\eta|<0.35$. In Fig.~\ref{fig:res_ppg216_v2_ff} the pseudorapidity 
acceptance of the FVTXS-CNT-FVTXN combination is listed as 
$-3.0<\eta<-1.0, |\eta|<0.35, 1.0<\eta<3.0$.  While the full FVTX 
acceptance for clusters is $1.0<|\eta|<3.0$, this analysis predominantly 
uses tracks that are from $1.2<|\eta|<2.2$.

\begin{table}[hbt]
\caption{Pseudorapidity acceptances of individual detectors.}
\begin{ruledtabular}
\begin{tabular}{lll}
Detector & $\eta_{{\rm min}}$ & $\eta_{{\rm max}}$ \\
\hline
BBCS (tubes) & -3.9 & -3.1 \\
FVTXS (clusters) & -3.0 & -1.0 \\
FVTXS (tracks) & -2.2 & -1.2 \\
CNT (tracks) & -0.35 & 0.35 \\
FVTXN (clusters) & 1.0 & 3.0 \\
FVTXN (tracks) & 1.2 & 2.2 \\
\end{tabular}
\end{ruledtabular}
\label{tab:detaci}
\caption{Pseudorapidity acceptances of two-detector combinations.}
\begin{ruledtabular}
\begin{tabular}{lll}
Detector combination & $|\Delta\eta|_{{\rm min}}$ & $|\left<\Delta\eta\right>|$ \\
\hline
BBCS-FVTXS (tubes-clusters) & 0.1 & 1.8 \\
BBCS-FVTXS (tubes-tracks) & 0.9 & 1.8 \\
BBCS-CNT (tubes-tracks) & 2.75 & 3.5 \\
FVTXS-CNT (clusters-tracks) & 0.65 & 2.0 \\
FVTXS-CNT (tracks-tracks) & 0.85 & 2.0 \\
FVTXN-CNT (clusters-tracks) & 0.65 & 2.0 \\
FVTXN-CNT (tracks-tracks) & 0.85 & 2.0 \\
FVTXS-FVTXN (clusters-clusters) & 2.0 & 3.4 \\
FVTXS-FVTXN (tracks-tracks) & 2.4 & 3.4 \\
\end{tabular}
\end{ruledtabular}
\label{tab:detacc}
\end{table}

\begin{figure}[hbt]
\includegraphics[width=1.0\linewidth]{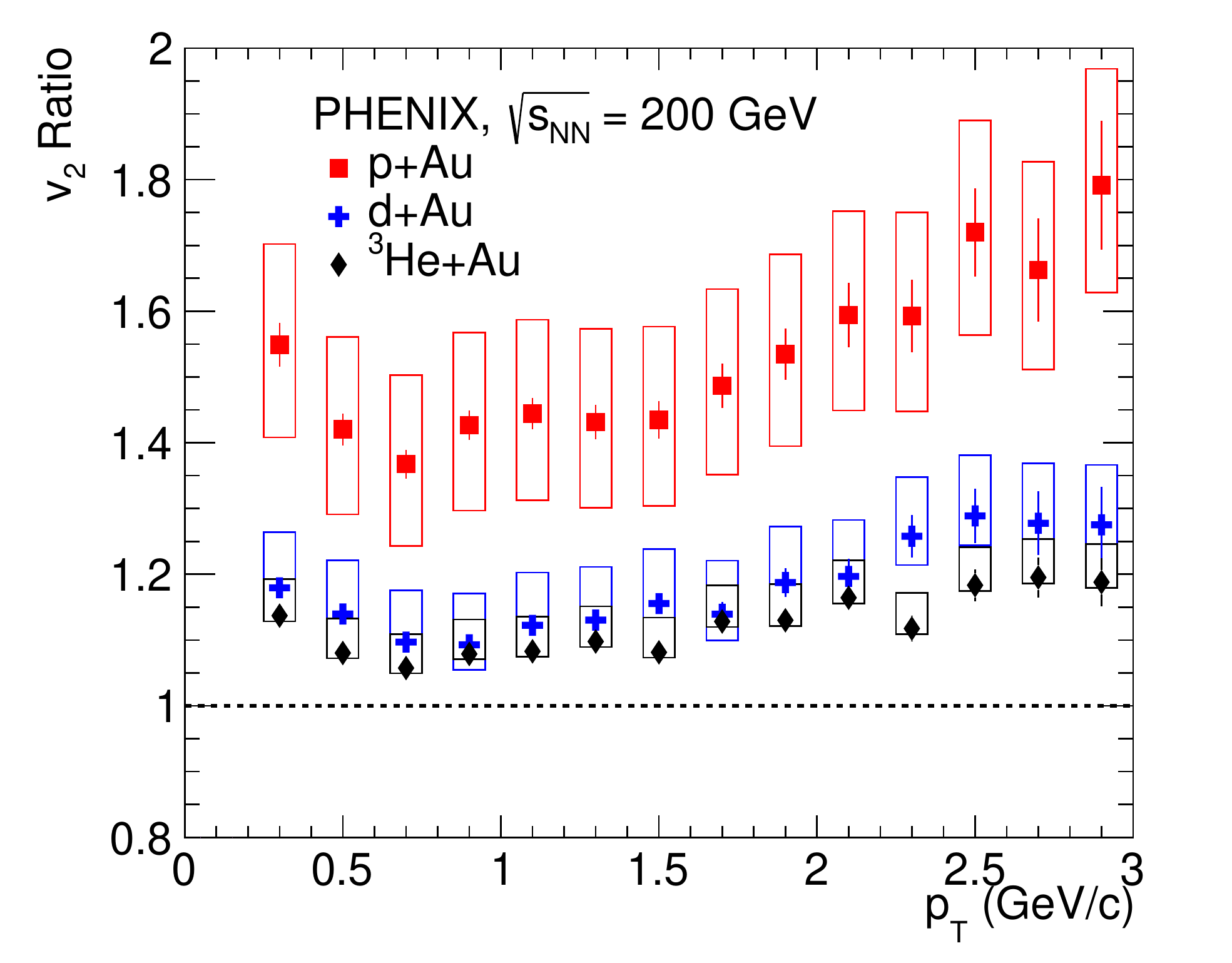}
\vspace{-0.4cm}
\caption{
Ratio of $v_2$ in the FVTXS-CNT-FVTXN combination (numerator) to $v_2$ 
in the BBCS-FVTXS-CNT combination (denominator) as a function of \pt for 
\pau (red squares), \dau (blue crosses), and \heau (black diamonds).
}
\label{fig:v2_ratio}
\end{figure}

Figure~\ref{fig:res_ppg216_v2_ff} shows the elliptic $v_{2}$ 
coefficients as a function of \pt from the \ttpc method utilizing the 
three-detector combination FVTXS-CNT-FVTXN.  The results for the most 
central 0\%--5\% events are shown for all three collision geometries with 
statistical uncertainties as vertical lines and systematic uncertainties 
as open boxes.  The systematic uncertainties have a high degree of 
point-to-point correlation. For comparison, the \ttpc values from the 
BBCS-FVTXS-CNT combination are shown.  Shown in 
Figure~\ref{fig:v2_ratio} is the ratio of the $v_2$ in the 
FVTXS-CNT-FVTXN combination to the $v_2$ in the BBCS-FVTXS-CNT 
combination.  We observe a modest 5\%--15\% difference in the \heau 
case, growing to a 10\%--20\% difference in the \dau case, and then a 
rather large 35\%--80\% difference in the \pau case.  Qualitatively this 
difference could result from substantially larger nonflow contributions 
in the FVTXS-CNT-FVTXN combination, and this would be expected to be 
largest in the \pau system which has the smallest multiplicity as well 
as the lowest expected elliptic flow itself.  Nonflow effects are also 
expected to play a larger role at larger $p_T$, and we observe a modest 
rise in the ratios with $p_T$.

\begin{figure*}[hbt]
\begin{minipage}{0.965\linewidth}
\includegraphics[width=0.32\linewidth]{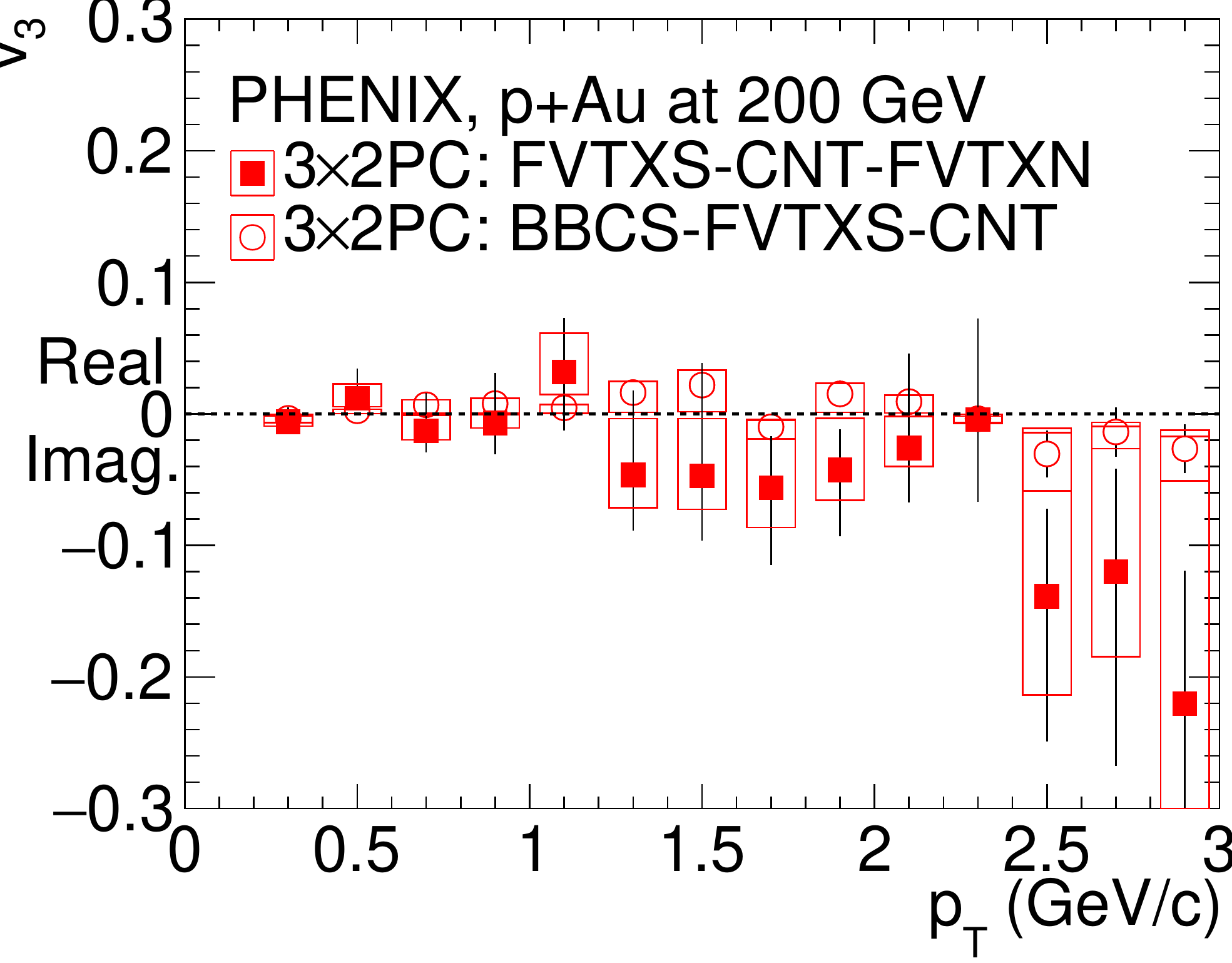}
\includegraphics[width=0.32\linewidth]{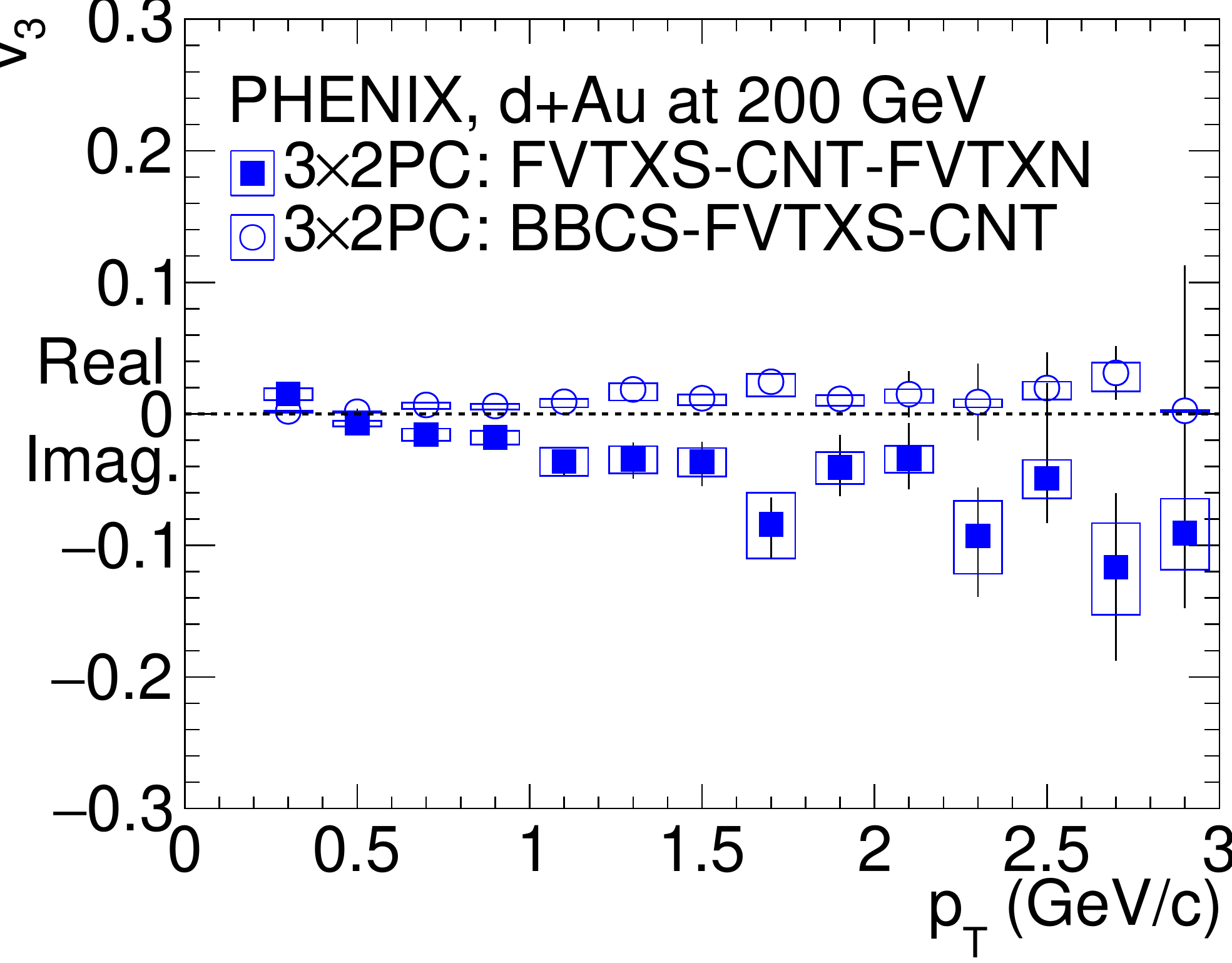}
\includegraphics[width=0.32\linewidth]{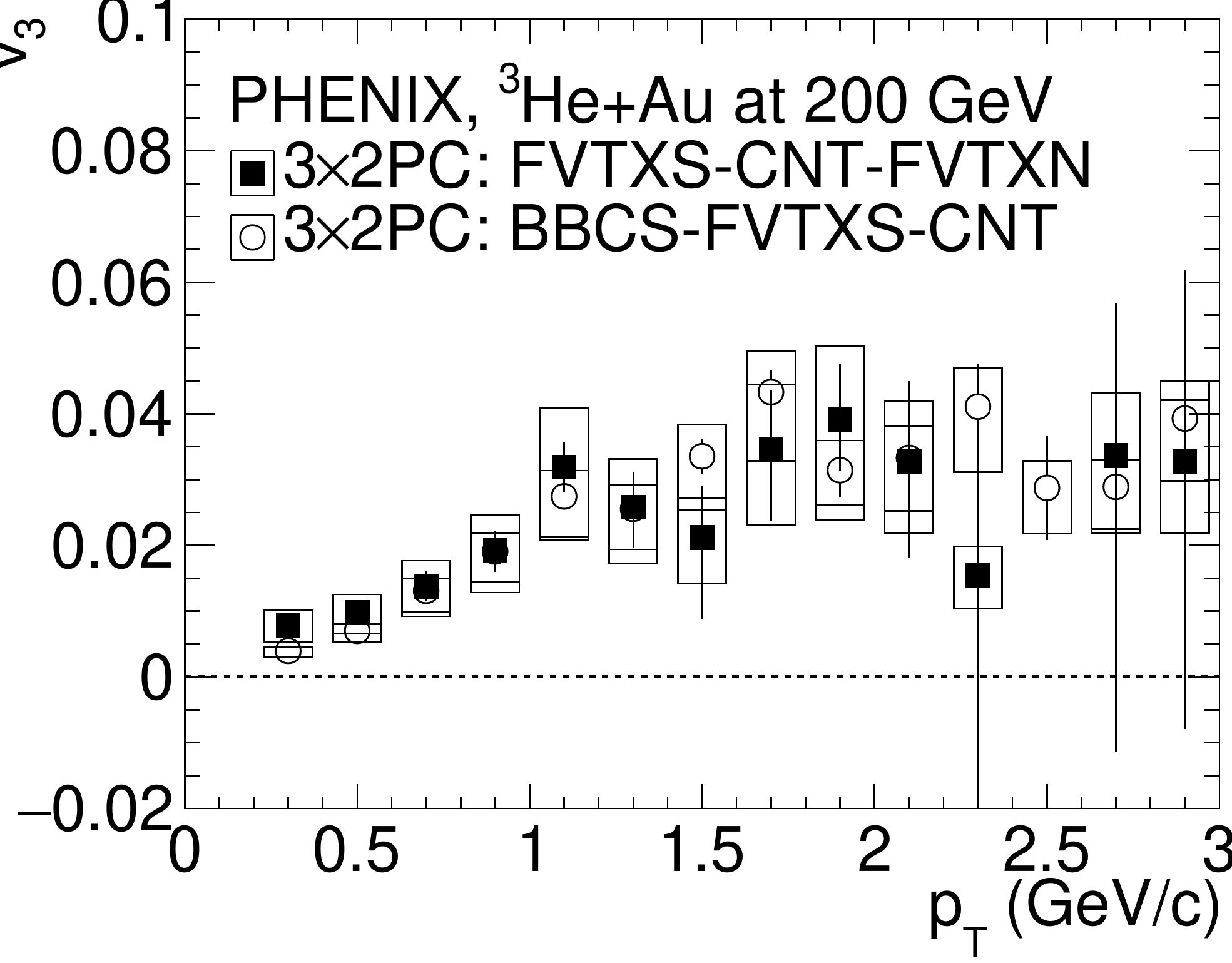}
\vspace{-0.4cm}
\caption{
The extracted $v_3$ coefficient as a function of \pt in 0\%--5\% central 
$p$$+$Au, $d$$+$Au, and $^3$He+Au collisions from 
the \ttpc method using the FVTXS-CNT-FVTXN detector combination are 
shown as solid squares.  For comparison we also show the previously 
plotted results from the BBCS-FVTXS-CNT combination as open circles.
}
\label{fig:res_ppg216_v3_ff}
\end{minipage}
\begin{minipage}{0.965\linewidth}
    \includegraphics[width=0.998\linewidth]{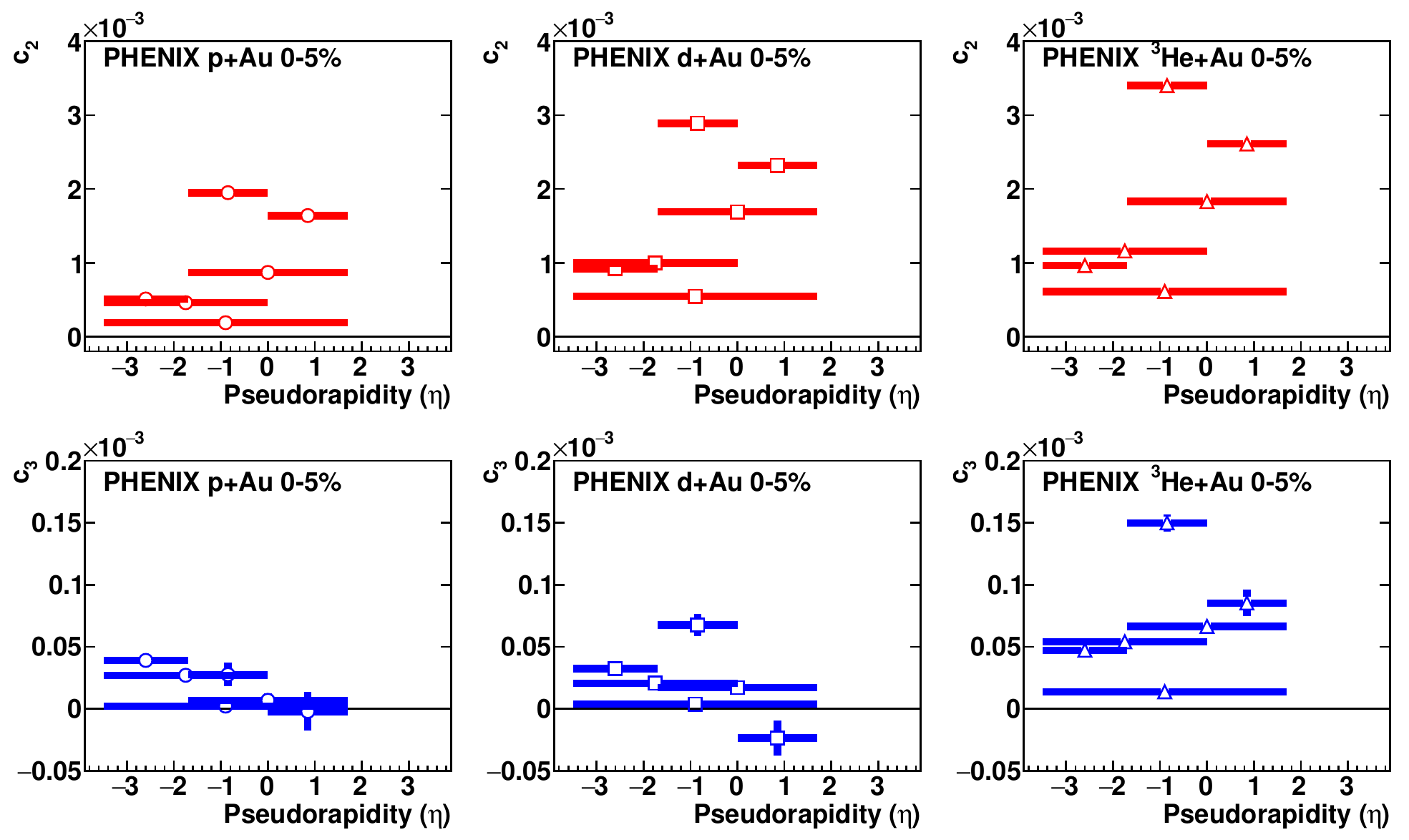}
\vspace{-0.4cm}
\caption{
Two-particle correlation $c_2$ and $c_3$ 
coefficients from 0\%--5\% central \pau, \dau, \heau collisions.  
The markers are located at the pseudorapidity average 
from the two detectors (i.e. $(\eta_{1}+\eta_{2})/2$) and the associated 
horizontal line extends between the two detectors (i.e. from $\eta_{1}$ 
to $\eta_{2}$).  The vertical bars indicate the statistical 
uncertainties.  See text for details.
}
    \label{fig:c2c3}
\end{minipage}
\end{figure*}

Figure~\ref{fig:res_ppg216_v3_ff} shows the third harmonic coefficient 
$v_{3}$ as a function of \pt from the \ttpc using the FVTXS-CNT-FVTXN 
method, shown in solid squares, and also the BBCS-FVTXS-CNT detector 
combination, shown in open circles, for comparison. The statistical 
uncertainties are shown as vertical lines and the systematic 
uncertainties as open boxes. In the case of \heau, the results agree for 
the two detector combination sets within uncertainties.  However, in the 
case of \pau and \dau, one of the $c_3$ coefficients is negative, and 
thus the mathematical calculation of $v_3$ results in an imaginary 
value. These imaginary values are shown along the negative $y$-axis in 
the figure.

These negative values of $c_3$ observed in the \pau and \dau systems are 
consistent with the observation that nonflow contributions in \pp 
collisions extrapolated to these systems drive $c_3$ towards negative 
values. This effect is consistent with nonflow dominance in the 
FVTXS-CNT-FVTXN result. It is striking how much larger the effect is in 
\pau and \dau compared to negligible in \heau.  Also, the 
difference in potential nonflow in the $v_2$ shown above is quite 
different between the systems and will also depend on the real 
triangular flow in these different geometries.

From the two-particle correlations and extracted coefficients (tabulated in 
the supplemental material~\cite{supp_mat}) one can examine the patterns 
between the two-particle kinematics and between collision systems.  Access 
to the full suite of Fourier coefficients is critical to enable future 
analysis techniques to be applied and comparison with new theoretical tools 
that might more fully incorporate flow, nonflow, and longitudinal dynamics. 
Figure~\ref{fig:c2c3} shows the $c_2$ and $c_3$ coefficients from 0\%--5\% 
central \pau, \dau, \heau collisions from left to right.  The markers are 
located at the pseudorapidity average from the two detectors (i.e. 
$(\eta_{1}+\eta_{2})/2$) and the associated horizontal line extends between 
the two detectors (i.e. from $\eta_{1}$ to $\eta_{2}$).  Correlations 
involving tracks in the CNT (e.g. where one of the horizontal line end 
points is at $\eta=0$) are for the inclusive range in $0.2<p_{T}<3.0~{\rm 
GeV}/c$.  As discussed in the supplemental material~\cite{supp_mat}, the $c_{n}$ 
coefficients should not be viewed as strict physics quantities because the 
charge in the BBC and tracks in the FVTX are not corrected for variations in 
acceptance, efficiency, and backgrounds, all of which can vary between 
running periods.

Starting with the $c_2$ values, one observes significant variation 
amongst the values from the different detector combinations used for the 
two-particle correlations.  This arises naturally from the 
pseudorapidity dependence of the flow $v_2$ itself, and also from the 
different $p_T$ coverage of the different detectors and different 
contributions from background, particularly in the BBCS.  Overall one 
observes that the relative ordering of $c_2$ values from different 
combinations is qualitatively similar for the three collision systems, 
with the dominant feature that all of the \pau values are lower.

For the $c_3$ values, the ordering of the detector combinations in \heau 
collisions is qualitatively similar to that of the $c_2$ values.  In 
striking contrast, all of the $c_3$ coefficients (so all detector 
combinations) are significantly lower in \pau and \dau compared with 
\heau.  This means that the conclusion of lower triangular flow in \pau 
and \dau is independent of any single detector used in the two-particle 
detector combination, i.e. it is seen in all combinations.  In 
particular, $c_3$ values where the FVTXN is utilized, i.e. where the 
horizontal line extends to $\eta=+1.7$, are very low and in some cases 
actually negative, though with large statistical uncertainties.  It is 
the negative value for $c_3$ between the CNT-FVTXN that results in the 
imaginary calculated $v_3$ in the FVTXS-CNT-FVTXN combination for \pau 
and \dau systems.  Noting that the multiplicity is lowest in these 
systems at forward rapidity, i.e. the proton or deuteron-going 
direction, and the $v_3$ may be the smallest, the explanation may be 
from a large nonflow contribution toward negative values of $c_3$.


\section{Summary}

In this paper we have presented an independent analysis of the flow 
coefficients $v_2$ and $v_3$ as a function of \pt in 0\%--5\% central 
\pau, \dau, and \heau collisions at \sqsn = 200 GeV using the \ttpc 
method.  The results are in excellent agreement with the published 
Nature Physics results from the PHENIX Collaboration using the event 
plane method~\cite{PHENIX:2018lia}. In addition, variations in the 
kinematic selection for the three detector combinations reveals an 
important role for nonflow and longitudinal decorrelations, 
particularly at forward rapidity, i.e. in the small projectile 
direction.  To support future analyses, this paper includes an archival 
documentation of correlation functions from \pp through \heau systems.



\vspace{-0.5cm}


\begin{acknowledgments}

We thank the staff of the Collider-Accelerator and Physics
Departments at Brookhaven National Laboratory and the staff of
the other PHENIX participating institutions for their vital
contributions.  
We acknowledge support from the Office of Nuclear Physics in the
Office of Science of the Department of Energy,
the National Science Foundation,
Abilene Christian University Research Council,
Research Foundation of SUNY, and
Dean of the College of Arts and Sciences, Vanderbilt University
(U.S.A),
Ministry of Education, Culture, Sports, Science, and Technology
and the Japan Society for the Promotion of Science (Japan),
Natural Science Foundation of China (People's Republic of China),
Croatian Science Foundation and
Ministry of Science and Education (Croatia),
Ministry of Education, Youth and Sports (Czech Republic),
Centre National de la Recherche Scientifique, Commissariat
{\`a} l'{\'E}nergie Atomique, and Institut National de Physique
Nucl{\'e}aire et de Physique des Particules (France),
J. Bolyai Research Scholarship, EFOP, the New National Excellence
Program ({\'U}NKP), NKFIH, and OTKA (Hungary),
Department of Atomic Energy and Department of Science and Technology
(India),
Israel Science Foundation (Israel),
Basic Science Research and SRC(CENuM) Programs through NRF
funded by the Ministry of Education and the Ministry of
Science and ICT (Korea).
Ministry of Education and Science, Russian Academy of Sciences,
Federal Agency of Atomic Energy (Russia),
VR and Wallenberg Foundation (Sweden),
the U.S. Civilian Research and Development Foundation for the
Independent States of the Former Soviet Union,
the Hungarian American Enterprise Scholarship Fund,
the US-Hungarian Fulbright Foundation,
and the US-Israel Binational Science Foundation.

\end{acknowledgments}

\vspace{-0.5cm}

\section*{APPENDIX: $v_n$ coefficients with two different subsets of 
FVTX acceptance}

We have repeated the analyses of \ttpc and dividing the FVTXS and FVTXN 
into halves, either selecting tracks with $1.2<|\eta|<1.7$ or 
$1.7<|\eta|<2.2$.  Thus, the detector combinations will have increased 
or decreased rapidity gaps from the default analysis.

\begin{figure*}
\begin{minipage}{0.99\linewidth}
\vspace{-0.5cm}
\includegraphics[width=0.32\textwidth]{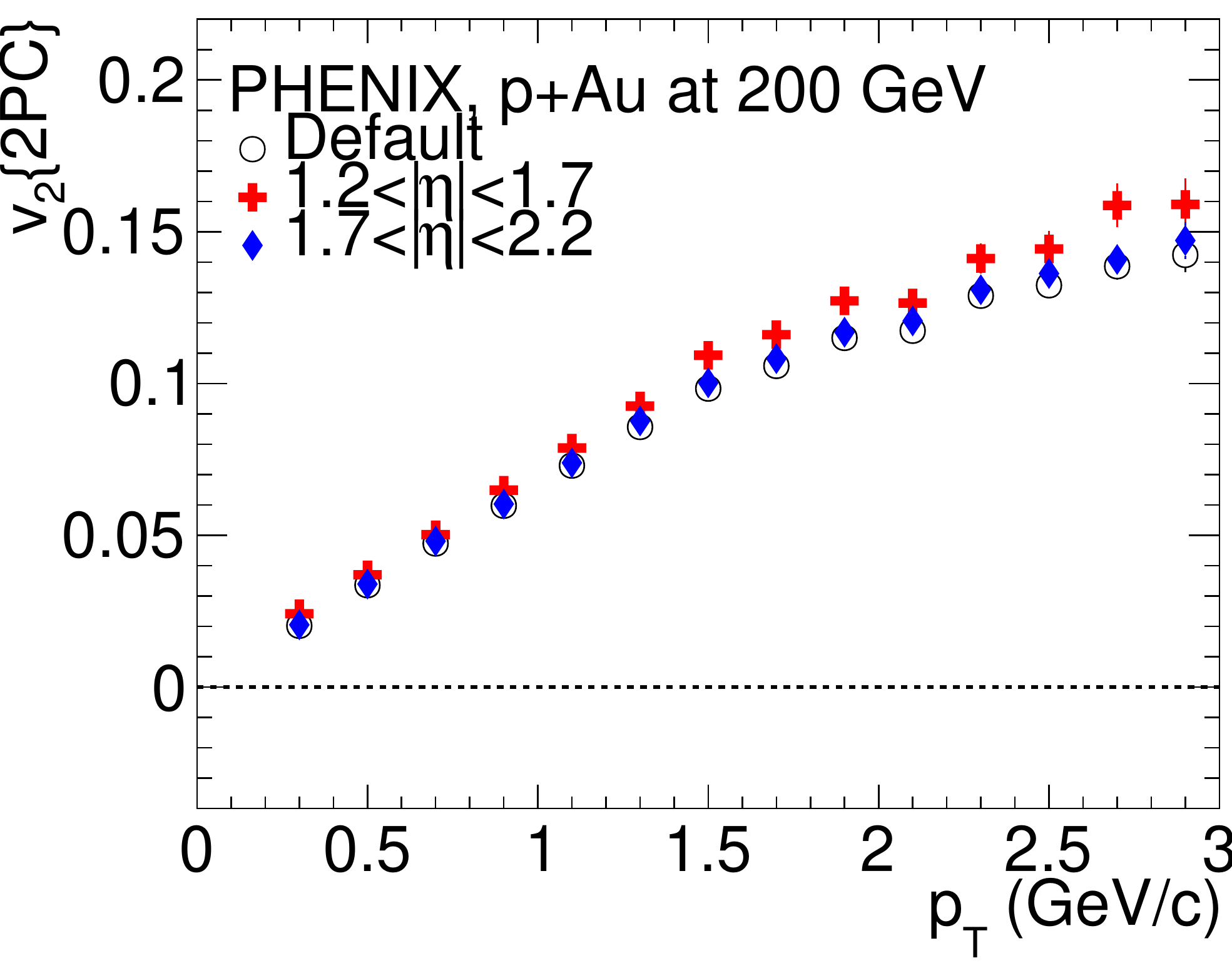}
\includegraphics[width=0.32\textwidth]{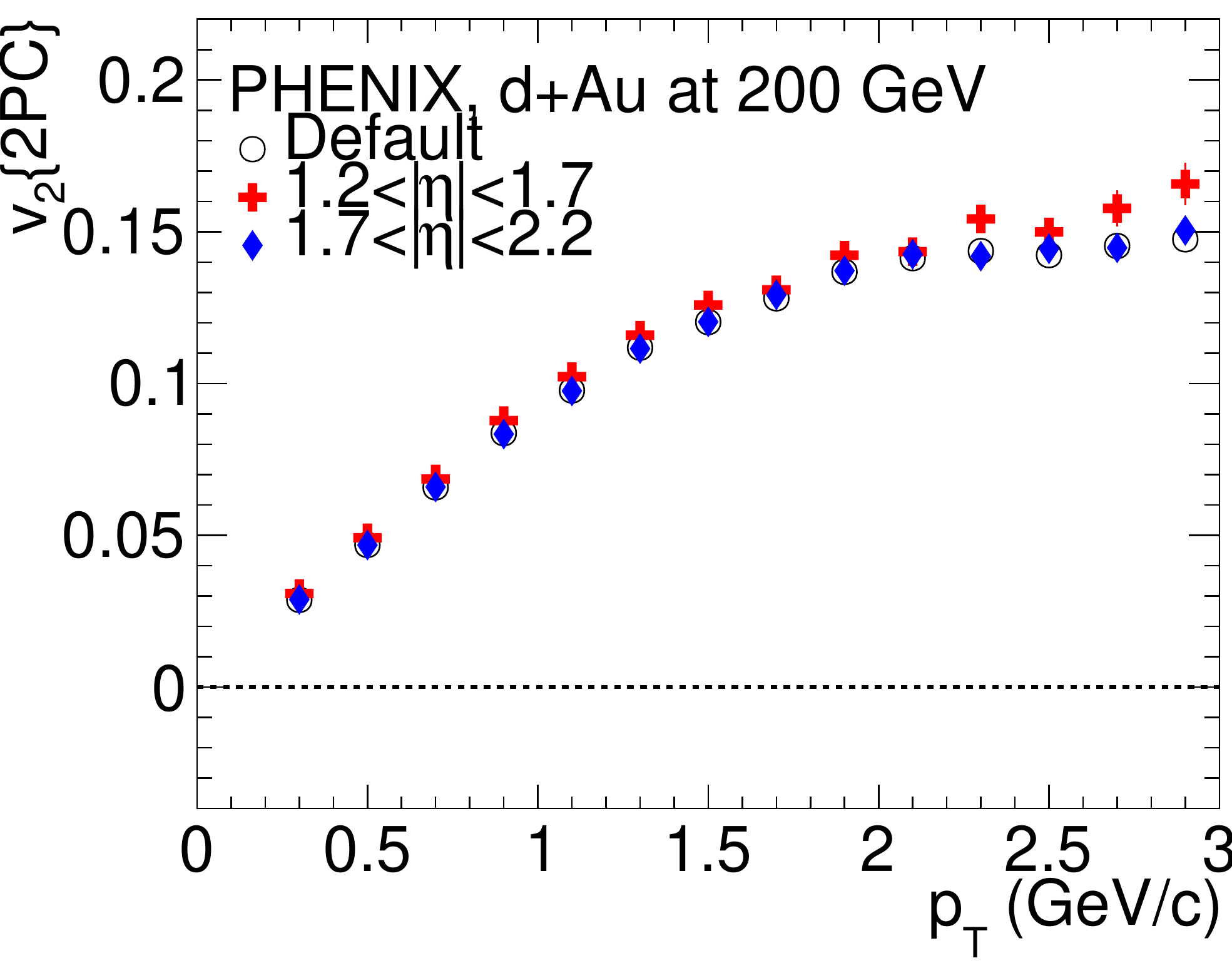}
\includegraphics[width=0.32\textwidth]{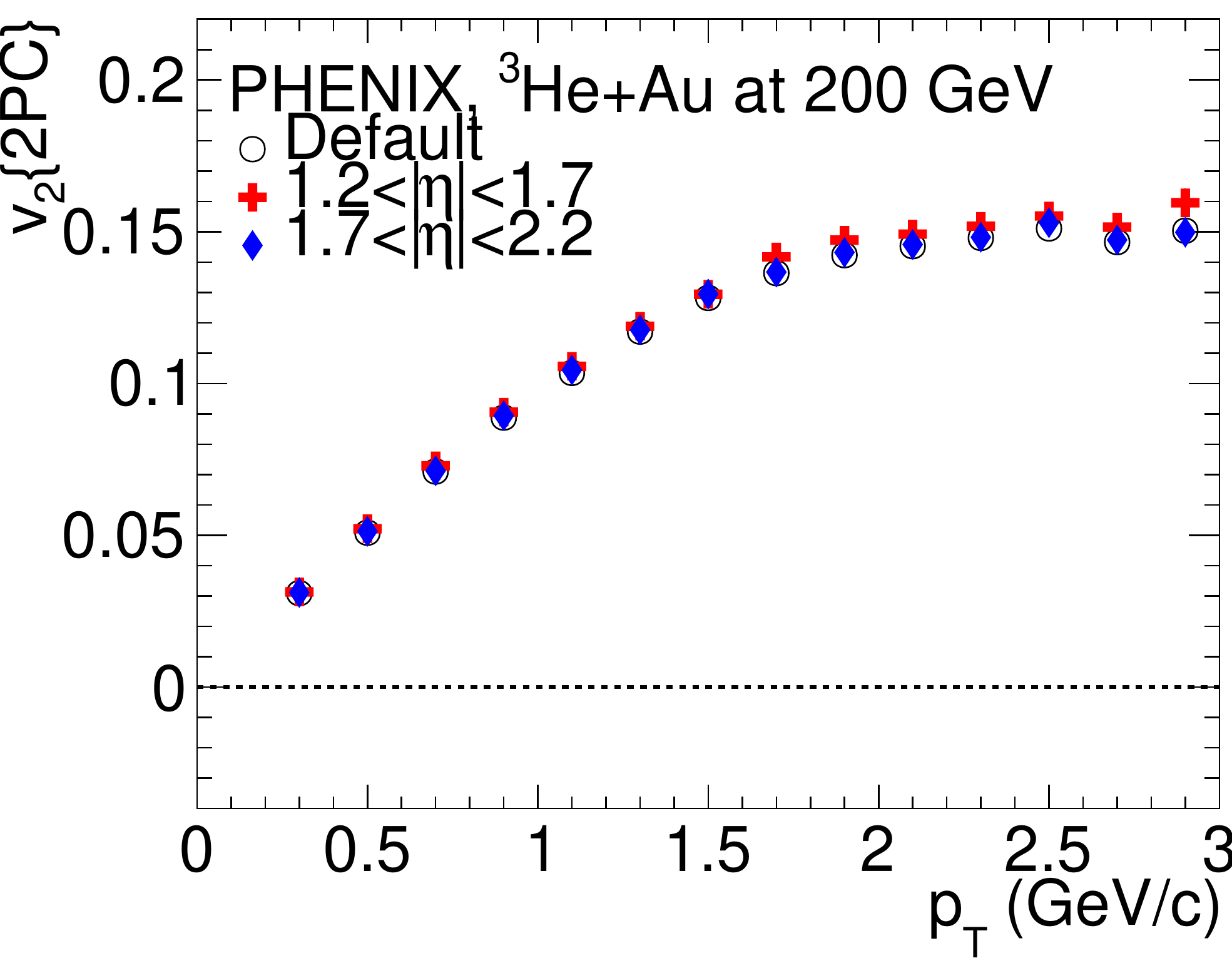}
\caption{
Calculated midrapidity $v_{2}$ as a function of \pt for 0\%--5\% central 
\pau, \dau, \heau collisions.  Shown are results from 
the BBCS-FVTXS-CNT combination including variations in the 
pseudorapidity selection of tracks in the FVTXS.
}
\label{fig:inout_v2bf}
\end{minipage}
\begin{minipage}{0.99\linewidth}
\vspace{0.5cm}
\includegraphics[width=0.32\textwidth]{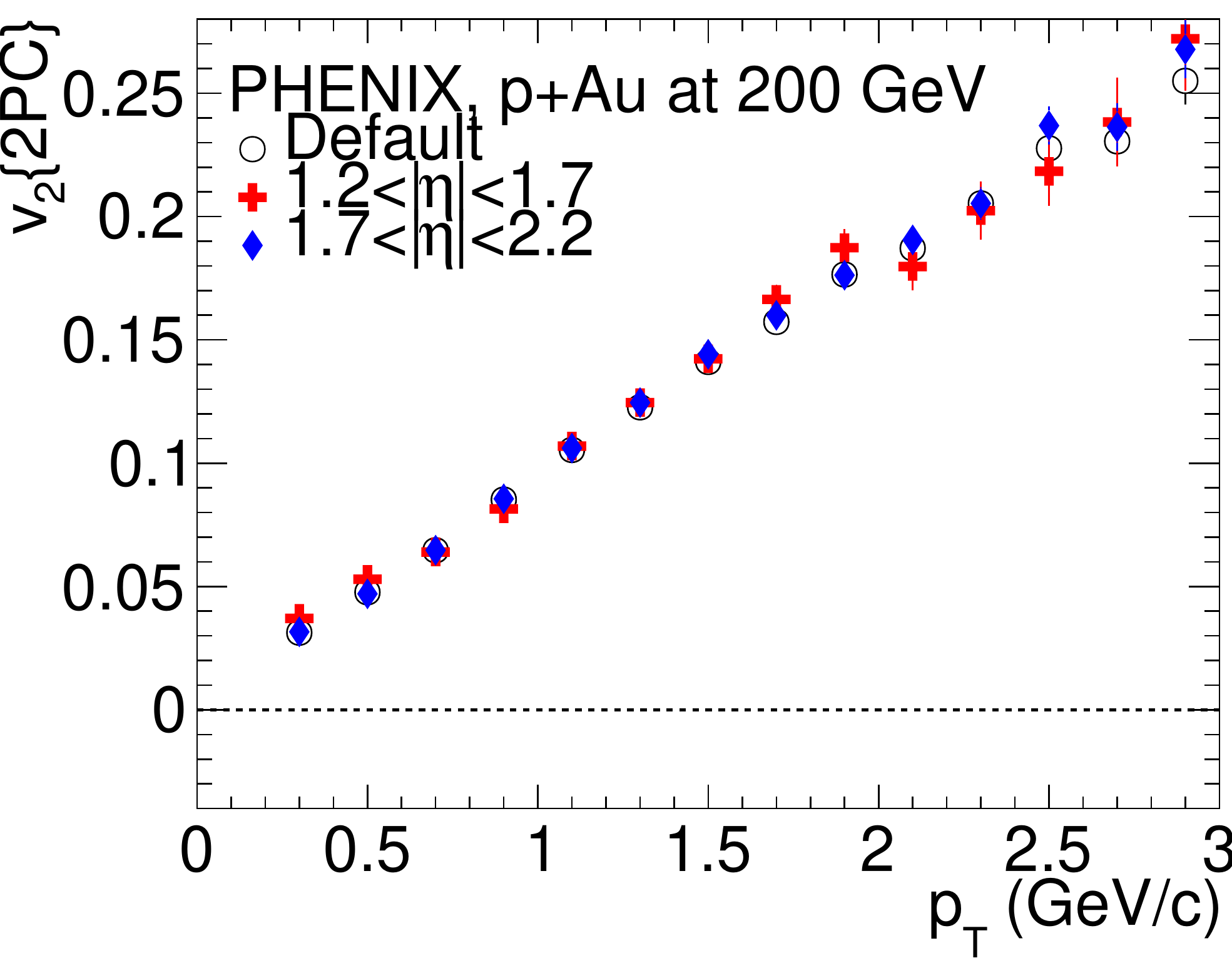}
\includegraphics[width=0.32\textwidth]{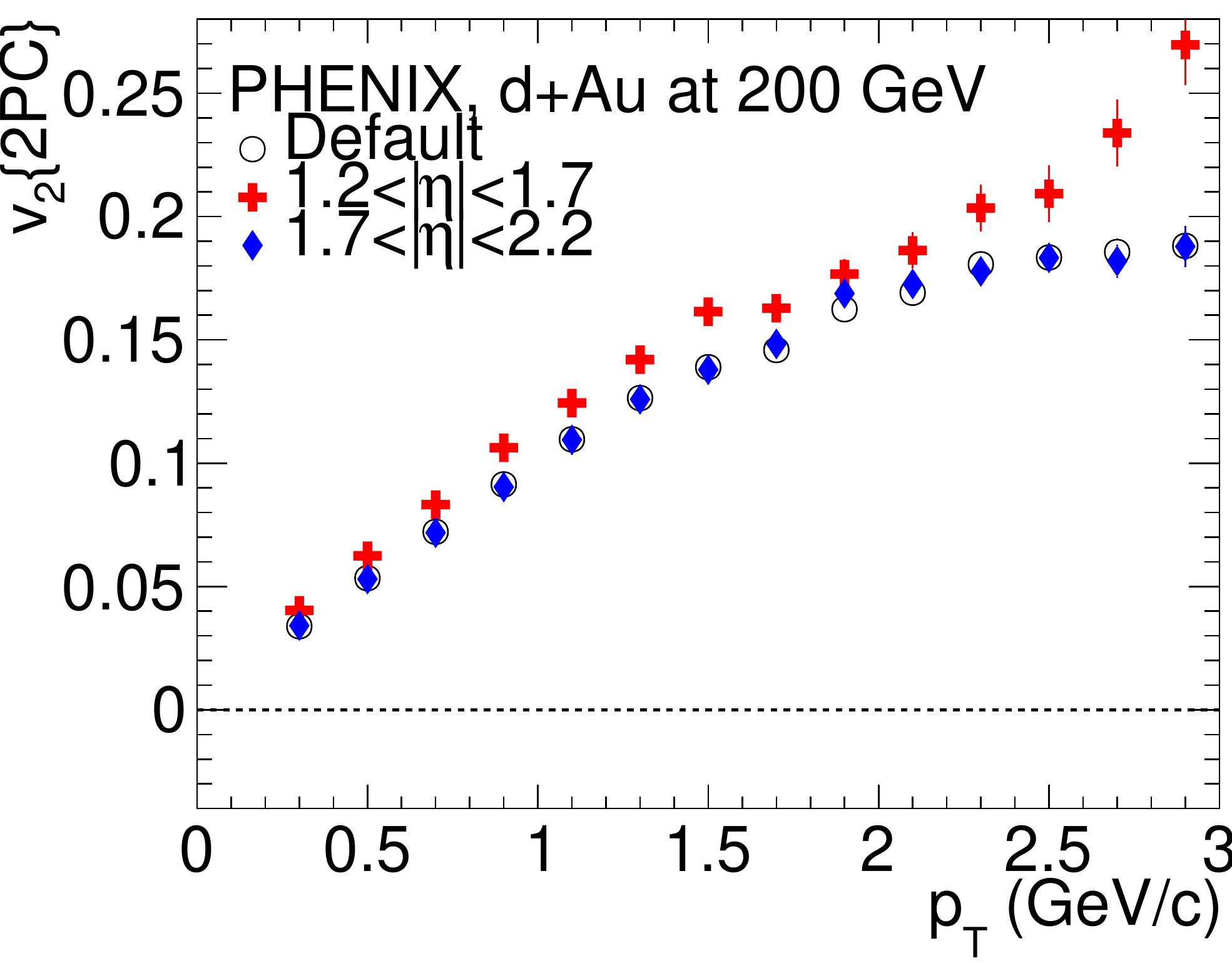}
\includegraphics[width=0.32\textwidth]{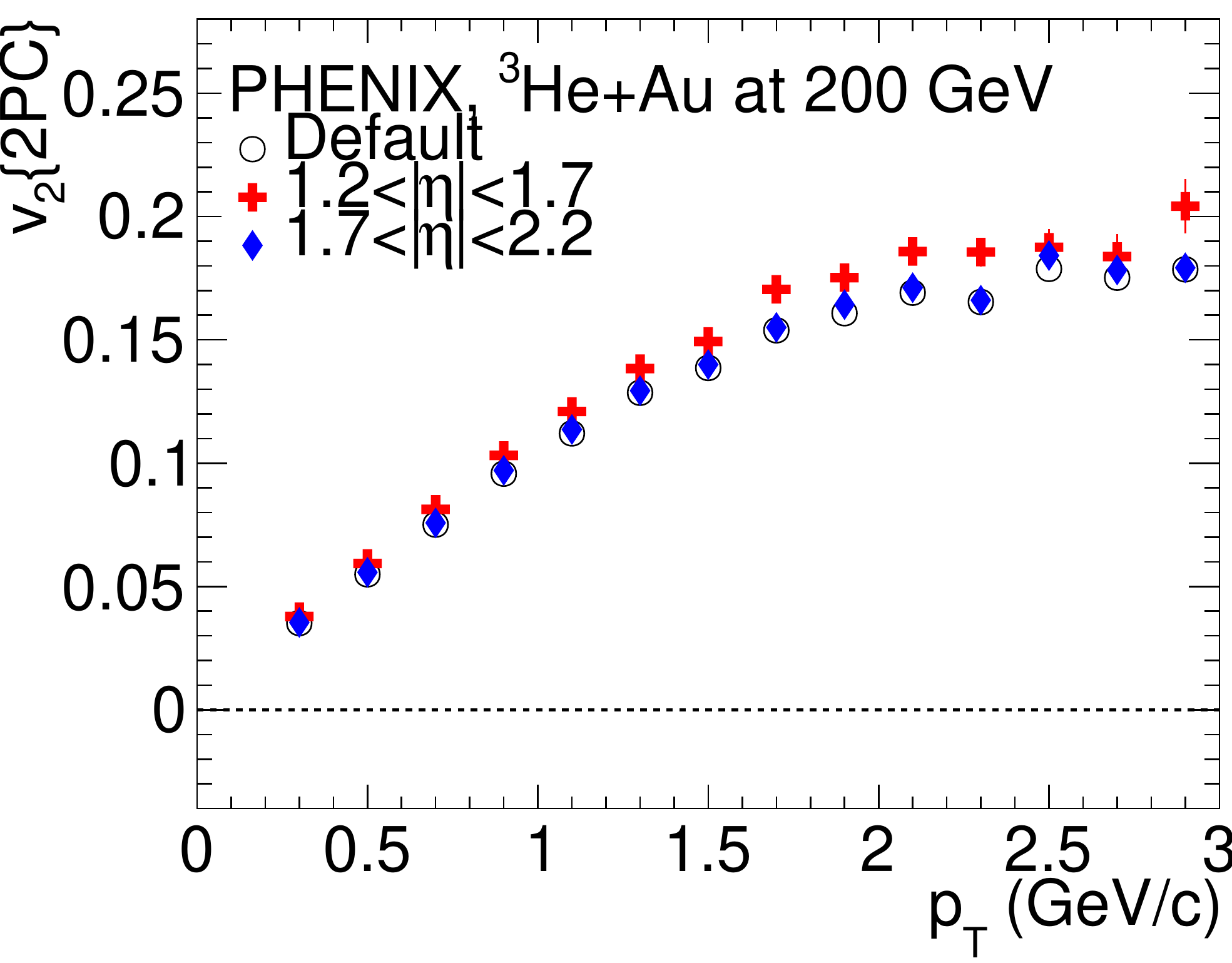}
\caption{
Calculated midrapidity $v_{2}$ as a function of \pt for 0\%--5\% central 
\pau, \dau, \heau collisions.  Shown are results from 
the FVTXS-CNT-FVTXN combination including variations in the 
pseudorapidity selection of tracks in the FVTXS and FVTXN.
}
\label{fig:inout_v2ff}
\end{minipage}
\begin{minipage}{0.99\linewidth}
\vspace{0.5cm}
\includegraphics[width=0.32\textwidth]{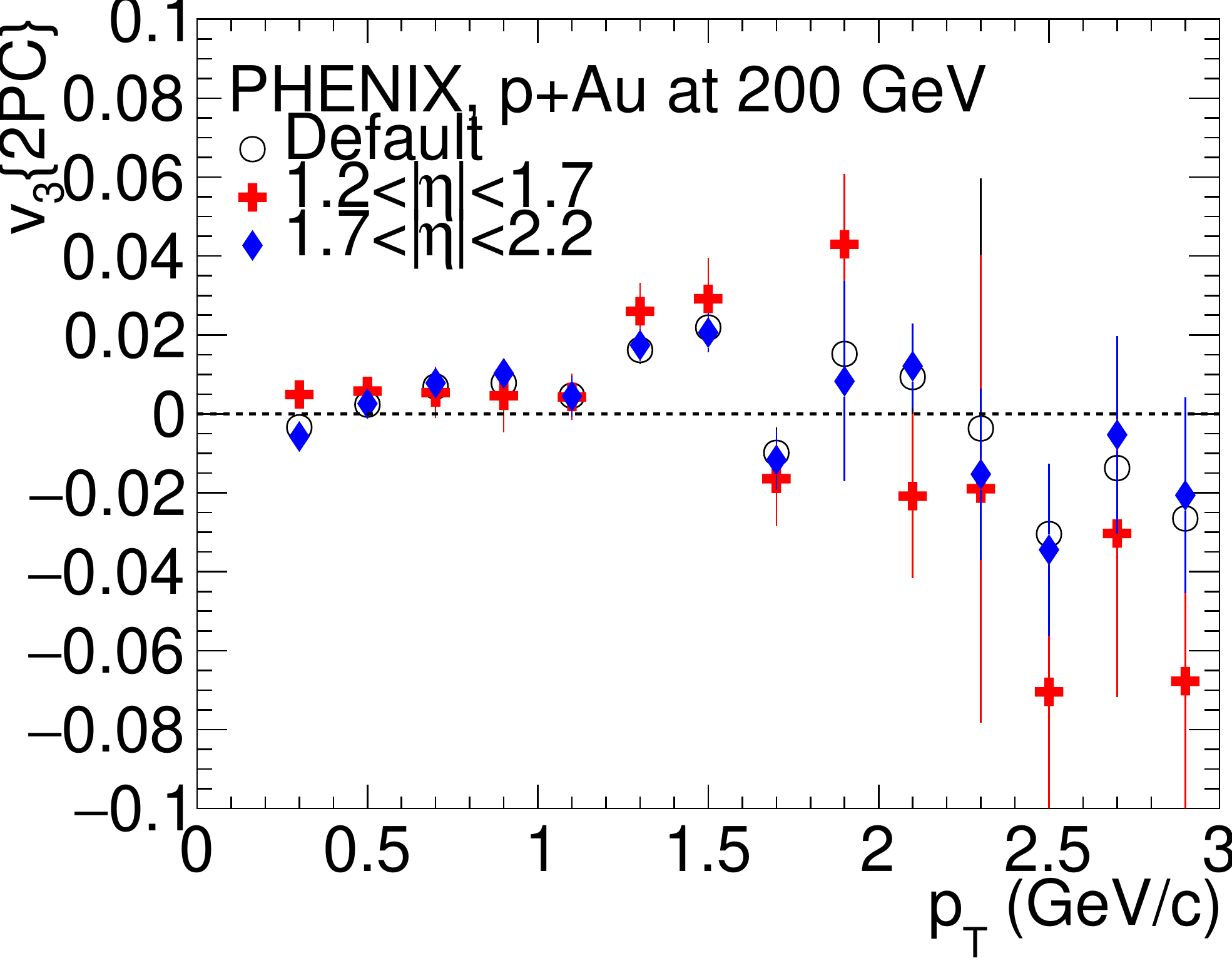}
\includegraphics[width=0.32\textwidth]{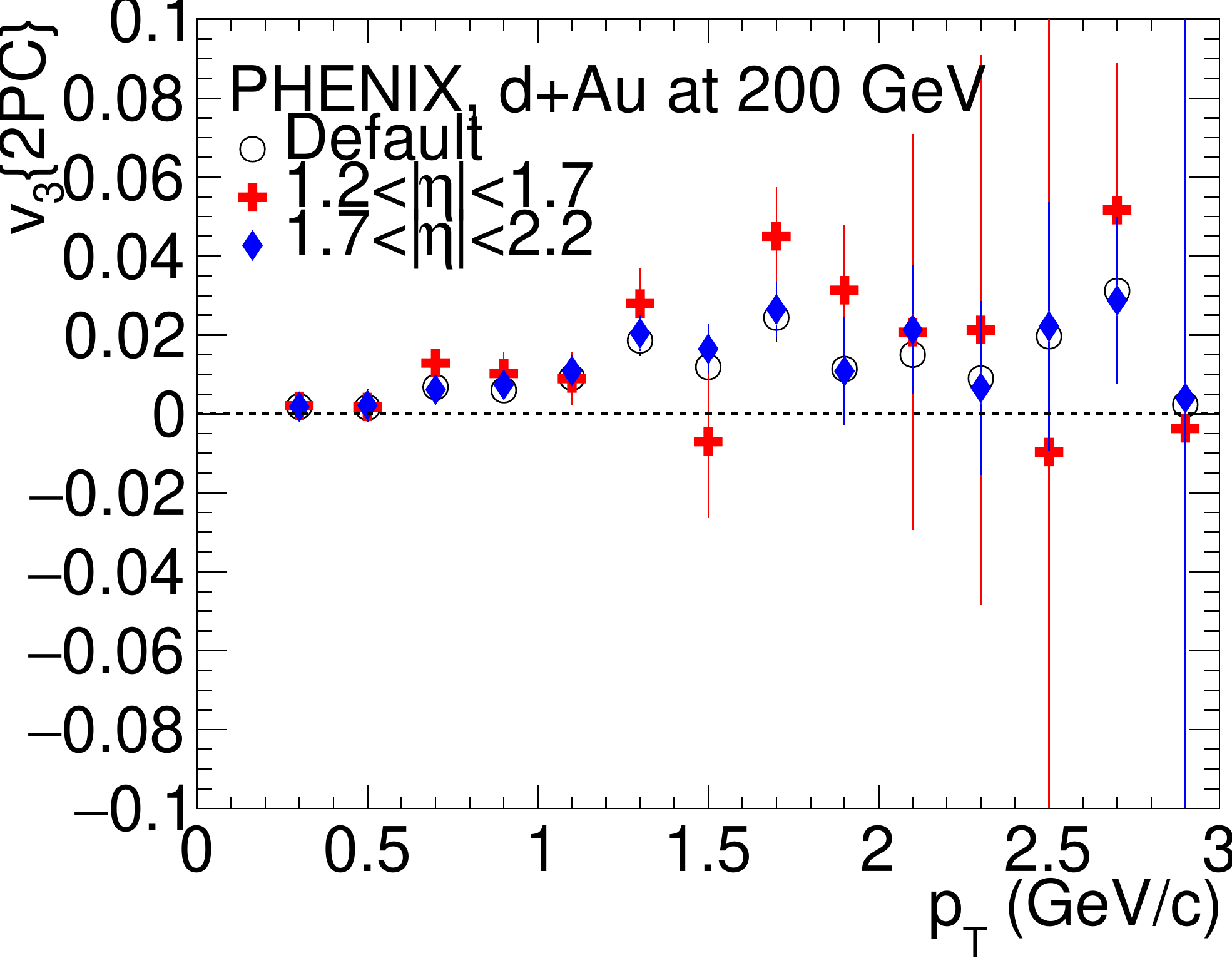}
\includegraphics[width=0.32\textwidth]{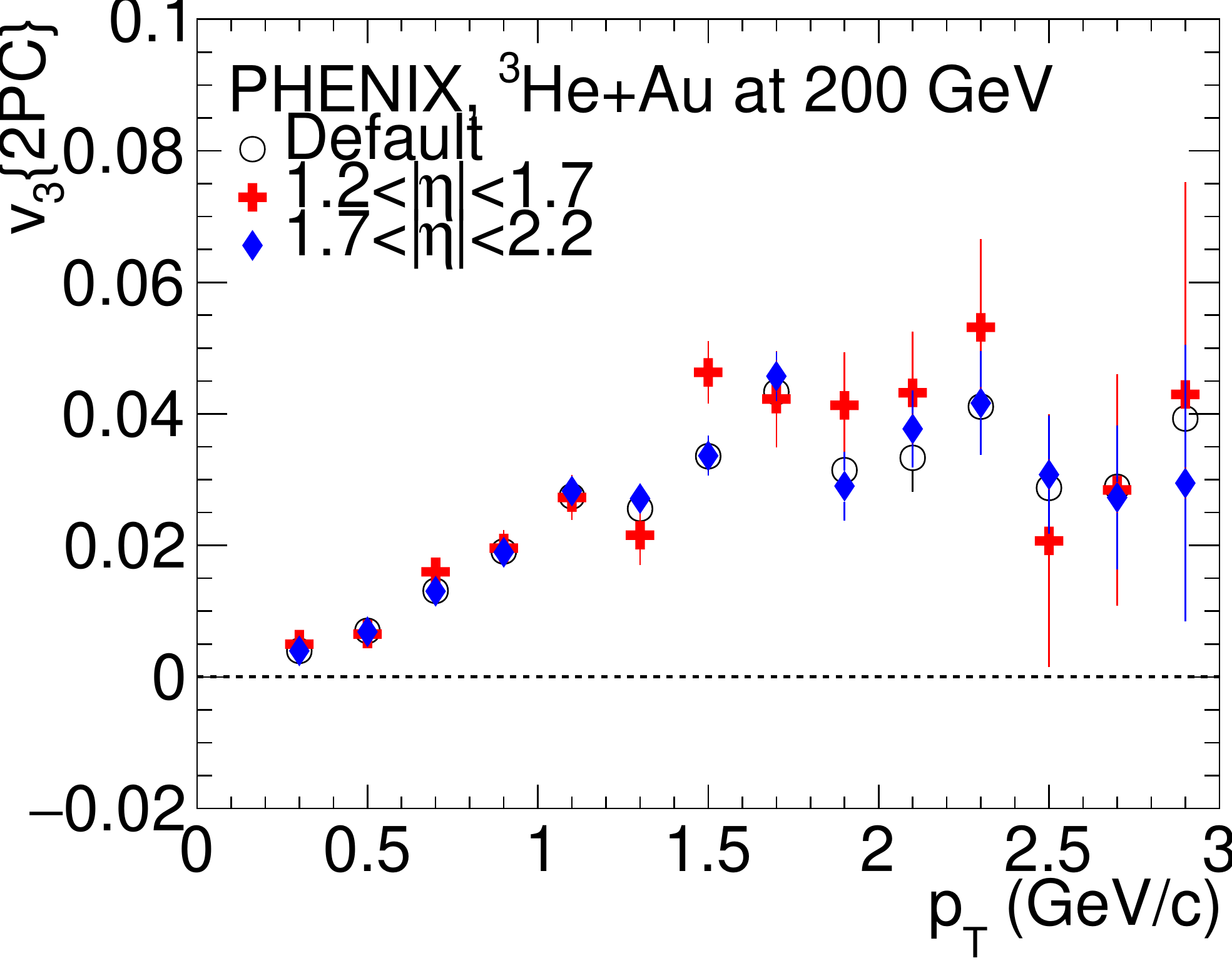}
\caption{
Calculated midrapidity $v_{3}$ as a function of \pt for 0\%--5\% central 
\pau, \dau, \heau collisions.  Shown are results from 
the BBCS-FVTXS-CNT combination including variations in the 
pseudorapidity selection of tracks in the FVTXS.
}
\label{fig:inout_v3bf}
\end{minipage}
\begin{minipage}{0.99\linewidth}
\vspace{0.5cm}
\includegraphics[width=0.32\textwidth]{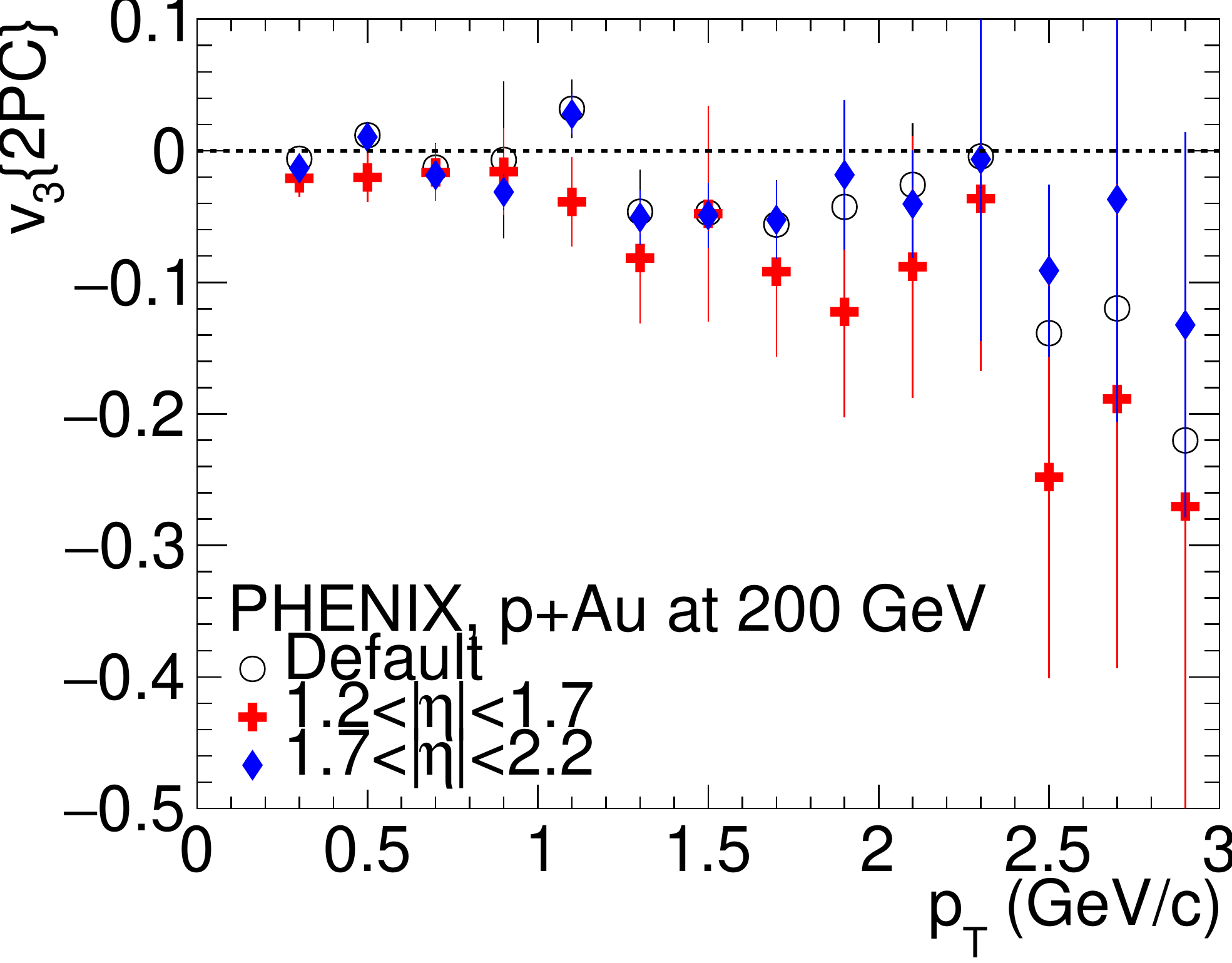}
\includegraphics[width=0.32\textwidth]{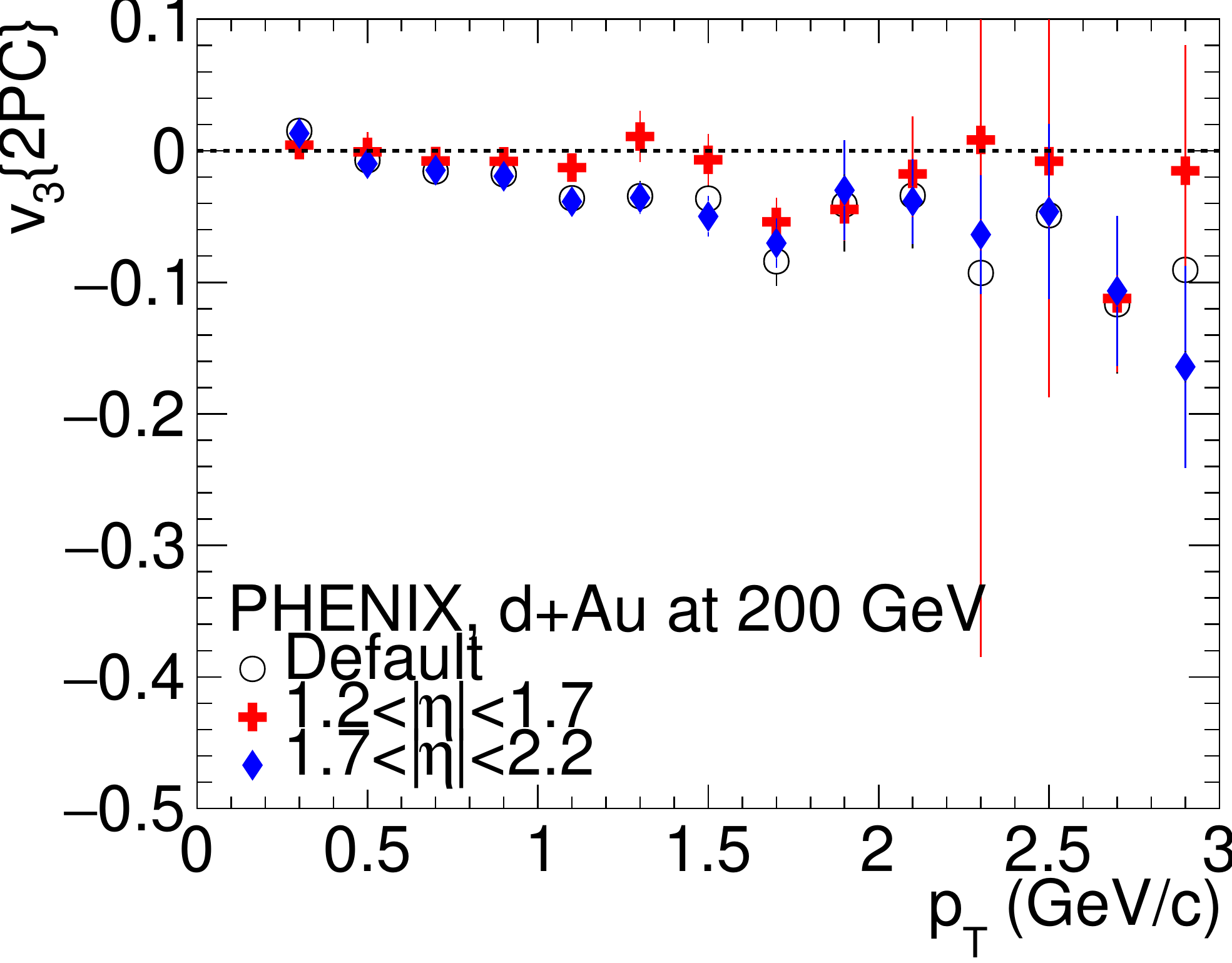}
\includegraphics[width=0.32\textwidth]{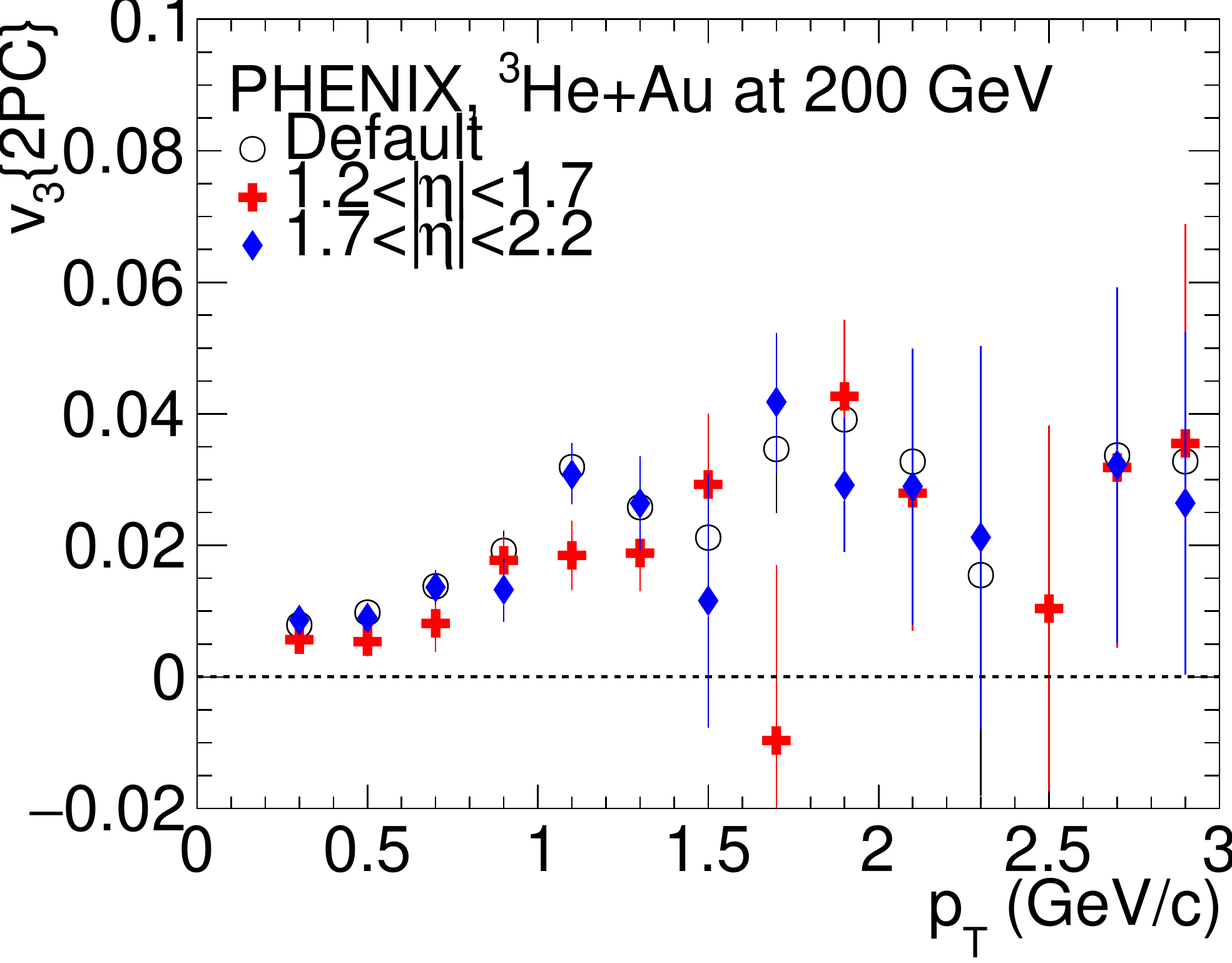}
\caption{
Calculated midrapidity $v_{3}$ as a function of \pt for 0\%--5\% central 
\pau, \dau, \heau collisions.  Shown are results from 
the FVTXS-CNT-FVTXN combination including variations in the 
pseudorapidity selection of tracks in the FVTXS and FVTXN.
}
\label{fig:inout_v3ff}
\end{minipage}
\end{figure*}

Figures~\ref{fig:inout_v2bf} and~\ref{fig:inout_v2ff} show the 
comparison of $v_{2}$ as a function of \pt at midrapidity using the two 
different FVTXN and FVTXS pseudorapidity ranges, in addition to the 
default use of the entire FVTX acceptance range.  In all cases, the 
differences are modest.  There is a general pattern that the $v_{2}$ 
calculated with the BBCS-FVTXS-CNT combination is slightly higher when 
using the FVTXS $1.2 < |\eta| < 1.7$ instead of FVTXS $1.7 < |\eta| < 
2.2$.  This may indicate a slight increase in nonflow contribution to 
the FVTXS-CNT correlation that dominates over a possible slight decrease 
in nonflow in the BBCS-FVTXS correlation.  A similar effect is seen in 
the $v_{2}$ values with the FVTXS-CNT-FVTXN combination, which again may 
related to slightly larger nonflow contributions due to all 
correlations having a smaller rapidity gap.

Figures~\ref{fig:inout_v3bf} and~\ref{fig:inout_v3ff} show the 
comparison of $v_{3}$ as a function of \pt at midrapidity using the two 
different FVTXN and FVTXS pseudorapidity ranges, in addition to the 
default use of the entire FVTX acceptance range. In all cases, the 
differences are modest, though with larger statistical uncertainties 
when splitting the FVTX acceptance.  The larger statistical 
uncertainties preclude any strong conclusions regarding a pattern with 
the different selections.

We do not include these differences as systematic uncertainties in the 
default $v_{2}$ and $v_{3}$ results as modest differences are expected.  
We can however rule out any large uncertainty from detector effects in 
the FVTX from the lower and higher rapidity acceptances.


%


\begin{thebibliography}{33}%
\makeatletter
\providecommand \@ifxundefined [1]{%
 \@ifx{#1\undefined}
}%
\providecommand \@ifnum [1]{%
 \ifnum #1\expandafter \@firstoftwo
 \else \expandafter \@secondoftwo
 \fi
}%
\providecommand \@ifx [1]{%
 \ifx #1\expandafter \@firstoftwo
 \else \expandafter \@secondoftwo
 \fi
}%
\providecommand \natexlab [1]{#1}%
\providecommand \enquote  [1]{``#1''}%
\providecommand \bibnamefont  [1]{#1}%
\providecommand \bibfnamefont [1]{#1}%
\providecommand \citenamefont [1]{#1}%
\providecommand \href@noop [0]{\@secondoftwo}%
\providecommand \href [0]{\begingroup \@sanitize@url \@href}%
\providecommand \@href[1]{\@@startlink{#1}\@@href}%
\providecommand \@@href[1]{\endgroup#1\@@endlink}%
\providecommand \@sanitize@url [0]{\catcode `\\12\catcode `\$12\catcode
  `\&12\catcode `\#12\catcode `\^12\catcode `\_12\catcode `\%12\relax}%
\providecommand \@@startlink[1]{}%
\providecommand \@@endlink[0]{}%
\providecommand \url  [0]{\begingroup\@sanitize@url \@url }%
\providecommand \@url [1]{\endgroup\@href {#1}{\urlprefix }}%
\providecommand \urlprefix  [0]{URL }%
\providecommand \Eprint [0]{\href }%
\providecommand \doibase [0]{https://doi.org/}%
\providecommand \selectlanguage [0]{\@gobble}%
\providecommand \bibinfo  [0]{\@secondoftwo}%
\providecommand \bibfield  [0]{\@secondoftwo}%
\providecommand \translation [1]{[#1]}%
\providecommand \BibitemOpen [0]{}%
\providecommand \bibitemStop [0]{}%
\providecommand \bibitemNoStop [0]{.\EOS\space}%
\providecommand \EOS [0]{\spacefactor3000\relax}%
\providecommand \BibitemShut  [1]{\csname bibitem#1\endcsname}%
\let\auto@bib@innerbib\@empty
\bibitem [{\citenamefont {Busza}\ \emph {et~al.}(2018)\citenamefont {Busza},
  \citenamefont {Rajagopal},\ and\ \citenamefont {van~der
  Schee}}]{Busza:2018rrf}%
  \BibitemOpen
  \bibfield  {author} {\bibinfo {author} {\bibfnamefont {W.}~\bibnamefont
  {Busza}}, \bibinfo {author} {\bibfnamefont {K.}~\bibnamefont {Rajagopal}},\
  and\ \bibinfo {author} {\bibfnamefont {W.}~\bibnamefont {van~der Schee}},\
  }\bibfield  {title} {\bibinfo {title} {{Heavy Ion Collisions: The Big
  Picture, and the Big Questions}},\ }\href
  {https://doi.org/10.1146/annurev-nucl-101917-020852} {\bibfield  {journal}
  {\bibinfo  {journal} {Ann. Rev. Nucl. Part. Sci.}\ }\textbf {\bibinfo
  {volume} {68}},\ \bibinfo {pages} {339} (\bibinfo {year} {2018})}\BibitemShut
  {NoStop}%
\bibitem [{\citenamefont {Muller}\ \emph {et~al.}(2012)\citenamefont {Muller},
  \citenamefont {Schukraft},\ and\ \citenamefont {Wyslouch}}]{Muller:2012zq}%
  \BibitemOpen
  \bibfield  {author} {\bibinfo {author} {\bibfnamefont {B.}~\bibnamefont
  {Muller}}, \bibinfo {author} {\bibfnamefont {J.}~\bibnamefont {Schukraft}},\
  and\ \bibinfo {author} {\bibfnamefont {B.}~\bibnamefont {Wyslouch}},\
  }\bibfield  {title} {\bibinfo {title} {{First Results from Pb+Pb collisions
  at the LHC}},\ }\href {https://doi.org/10.1146/annurev-nucl-102711-094910}
  {\bibfield  {journal} {\bibinfo  {journal} {Ann. Rev. Nucl. Part. Sci.}\
  }\textbf {\bibinfo {volume} {62}},\ \bibinfo {pages} {361} (\bibinfo {year}
  {2012})}\BibitemShut {NoStop}%
\bibitem [{\citenamefont {Muller}\ and\ \citenamefont
  {Nagle}(2006)}]{Muller:2006ee}%
  \BibitemOpen
  \bibfield  {author} {\bibinfo {author} {\bibfnamefont {B.}~\bibnamefont
  {Muller}}\ and\ \bibinfo {author} {\bibfnamefont {J.~L.}\ \bibnamefont
  {Nagle}},\ }\bibfield  {title} {\bibinfo {title} {{Results from the
  relativistic heavy ion collider}},\ }\href
  {https://doi.org/10.1146/annurev.nucl.56.080805.140556} {\bibfield  {journal}
  {\bibinfo  {journal} {Ann. Rev. Nucl. Part. Sci.}\ }\textbf {\bibinfo
  {volume} {56}},\ \bibinfo {pages} {93} (\bibinfo {year} {2006})}\BibitemShut
  {NoStop}%
\bibitem [{\citenamefont {Nagle}\ and\ \citenamefont
  {Zajc}(2018)}]{Nagle:2018nvi}%
  \BibitemOpen
  \bibfield  {author} {\bibinfo {author} {\bibfnamefont {J.~L.}\ \bibnamefont
  {Nagle}}\ and\ \bibinfo {author} {\bibfnamefont {W.~A.}\ \bibnamefont
  {Zajc}},\ }\bibfield  {title} {\bibinfo {title} {{Small System Collectivity
  in Relativistic Hadronic and Nuclear Collisions}},\ }\href
  {https://doi.org/10.1146/annurev-nucl-101916-123209} {\bibfield  {journal}
  {\bibinfo  {journal} {Ann. Rev. Nucl. Part. Sci.}\ }\textbf {\bibinfo
  {volume} {68}},\ \bibinfo {pages} {211} (\bibinfo {year} {2018})}\BibitemShut
  {NoStop}%
\bibitem [{\citenamefont {Nagle}\ \emph {et~al.}(2014)\citenamefont {Nagle},
  \citenamefont {Adare}, \citenamefont {Beckman}, \citenamefont {Koblesky},
  \citenamefont {Orjuela~Koop}, \citenamefont {McGlinchey}, \citenamefont
  {Romatschke}, \citenamefont {Carlson}, \citenamefont {Lynn},\ and\
  \citenamefont {McCumber}}]{Nagle:2013lja}%
  \BibitemOpen
  \bibfield  {author} {\bibinfo {author} {\bibfnamefont {J.~L.}\ \bibnamefont
  {Nagle}}, \bibinfo {author} {\bibfnamefont {A.}~\bibnamefont {Adare}},
  \bibinfo {author} {\bibfnamefont {S.}~\bibnamefont {Beckman}}, \bibinfo
  {author} {\bibfnamefont {T.}~\bibnamefont {Koblesky}}, \bibinfo {author}
  {\bibfnamefont {J.}~\bibnamefont {Orjuela~Koop}}, \bibinfo {author}
  {\bibfnamefont {D.}~\bibnamefont {McGlinchey}}, \bibinfo {author}
  {\bibfnamefont {P.}~\bibnamefont {Romatschke}}, \bibinfo {author}
  {\bibfnamefont {J.}~\bibnamefont {Carlson}}, \bibinfo {author} {\bibfnamefont
  {J.~E.}\ \bibnamefont {Lynn}},\ and\ \bibinfo {author} {\bibfnamefont
  {M.}~\bibnamefont {McCumber}},\ }\bibfield  {title} {\bibinfo {title}
  {{Exploiting Intrinsic Triangular Geometry in Relativistic He$^3$$+$Au
  Collisions to Disentangle Medium Properties}},\ }\href
  {https://doi.org/10.1103/PhysRevLett.113.112301} {\bibfield  {journal}
  {\bibinfo  {journal} {Phys. Rev. Lett.}\ }\textbf {\bibinfo {volume} {113}},\
  \bibinfo {pages} {112301} (\bibinfo {year} {2014})}\BibitemShut {NoStop}%
\bibitem [{\citenamefont {Adare}\ \emph
  {et~al.}(2013{\natexlab{a}})\citenamefont {Adare} \emph
  {et~al.}}]{Adare:2013esx}%
  \BibitemOpen
  \bibfield  {author} {\bibinfo {author} {\bibfnamefont {A.}~\bibnamefont
  {Adare}} \emph {et~al.} (\bibinfo {collaboration} {PHENIX Collaboration}),\
  }\bibfield  {title} {\bibinfo {title} {{Spectra and ratios of identified
  particles in Au+Au and $d$+Au collisions at $\sqrt{s_{NN}}=200$ GeV}},\
  }\href {https://doi.org/10.1103/PhysRevC.88.024906} {\bibfield  {journal}
  {\bibinfo  {journal} {Phys. Rev. C}\ }\textbf {\bibinfo {volume} {88}},\
  \bibinfo {pages} {024906} (\bibinfo {year} {2013}{\natexlab{a}})}\BibitemShut
  {NoStop}%
\bibitem [{\citenamefont {Adare}\ \emph
  {et~al.}(2018{\natexlab{a}})\citenamefont {Adare} \emph
  {et~al.}}]{Adare:2018toe}%
  \BibitemOpen
  \bibfield  {author} {\bibinfo {author} {\bibfnamefont {A.}~\bibnamefont
  {Adare}} \emph {et~al.} (\bibinfo {collaboration} {PHENIX Collaboration}),\
  }\bibfield  {title} {\bibinfo {title} {{Pseudorapidity Dependence of Particle
  Production and Elliptic Flow in Asymmetric Nuclear Collisions of $p+$Al,
  $p+$Au, $d+$Au, and $^{3}$He$+$Au at $\sqrt{s_{_{NN}}}=200$ GeV}},\ }\href
  {https://doi.org/10.1103/PhysRevLett.121.222301} {\bibfield  {journal}
  {\bibinfo  {journal} {Phys. Rev. Lett.}\ }\textbf {\bibinfo {volume} {121}},\
  \bibinfo {pages} {222301} (\bibinfo {year} {2018}{\natexlab{a}})}\BibitemShut
  {NoStop}%
\bibitem [{\citenamefont {Aidala}\ \emph
  {et~al.}(2017{\natexlab{a}})\citenamefont {Aidala} \emph
  {et~al.}}]{Aidala:2017pup}%
  \BibitemOpen
  \bibfield  {author} {\bibinfo {author} {\bibfnamefont {C.}~\bibnamefont
  {Aidala}} \emph {et~al.} (\bibinfo {collaboration} {PHENIX Collaboration}),\
  }\bibfield  {title} {\bibinfo {title} {{Measurements of azimuthal anisotropy
  and charged-particle multiplicity in $d+$Au collisions at
  $\sqrt{s_{_{NN}}}=$200, 62.4, 39, and 19.6 GeV}},\ }\href
  {https://doi.org/10.1103/PhysRevC.96.064905} {\bibfield  {journal} {\bibinfo
  {journal} {Phys. Rev. C}\ }\textbf {\bibinfo {volume} {96}},\ \bibinfo
  {pages} {064905} (\bibinfo {year} {2017}{\natexlab{a}})}\BibitemShut
  {NoStop}%
\bibitem [{\citenamefont {Aidala}\ \emph {et~al.}(2018)\citenamefont {Aidala}
  \emph {et~al.}}]{Aidala:2017ajz}%
  \BibitemOpen
  \bibfield  {author} {\bibinfo {author} {\bibfnamefont {C.}~\bibnamefont
  {Aidala}} \emph {et~al.} (\bibinfo {collaboration} {PHENIX Collaboration}),\
  }\bibfield  {title} {\bibinfo {title} {{Measurements of Multiparticle
  Correlations in $d+\mathrm{Au}$ Collisions at 200, 62.4, 39, and 19.6 GeV and
  $p+\mathrm{Au}$ Collisions at 200 GeV and Implications for Collective
  Behavior}},\ }\href {https://doi.org/10.1103/PhysRevLett.120.062302}
  {\bibfield  {journal} {\bibinfo  {journal} {Phys. Rev. Lett.}\ }\textbf
  {\bibinfo {volume} {120}},\ \bibinfo {pages} {062302} (\bibinfo {year}
  {2018})}\BibitemShut {NoStop}%
\bibitem [{\citenamefont {Aidala}\ \emph
  {et~al.}(2017{\natexlab{b}})\citenamefont {Aidala} \emph
  {et~al.}}]{Aidala:2016vgl}%
  \BibitemOpen
  \bibfield  {author} {\bibinfo {author} {\bibfnamefont {C.}~\bibnamefont
  {Aidala}} \emph {et~al.} (\bibinfo {collaboration} {PHENIX Collaboration}),\
  }\bibfield  {title} {\bibinfo {title} {{Measurement of long-range angular
  correlations and azimuthal anisotropies in high-multiplicity $p+$Au
  collisions at $\sqrt{s_{_{NN}}}=200$ GeV}},\ }\href
  {https://doi.org/10.1103/PhysRevC.95.034910} {\bibfield  {journal} {\bibinfo
  {journal} {Phys. Rev. C}\ }\textbf {\bibinfo {volume} {95}},\ \bibinfo
  {pages} {034910} (\bibinfo {year} {2017}{\natexlab{b}})}\BibitemShut
  {NoStop}%
\bibitem [{\citenamefont {Adare}\ \emph
  {et~al.}(2015{\natexlab{a}})\citenamefont {Adare} \emph
  {et~al.}}]{Adare:2015ctn}%
  \BibitemOpen
  \bibfield  {author} {\bibinfo {author} {\bibfnamefont {A.}~\bibnamefont
  {Adare}} \emph {et~al.} (\bibinfo {collaboration} {PHENIX Collaboration}),\
  }\bibfield  {title} {\bibinfo {title} {{Measurements of elliptic and
  triangular flow in high-multiplicity $^{3}$He$+$Au collisions at
  $\sqrt{s_{_{NN}}}=200$ GeV}},\ }\href
  {https://doi.org/10.1103/PhysRevLett.115.142301} {\bibfield  {journal}
  {\bibinfo  {journal} {Phys. Rev. Lett.}\ }\textbf {\bibinfo {volume} {115}},\
  \bibinfo {pages} {142301} (\bibinfo {year} {2015}{\natexlab{a}})}\BibitemShut
  {NoStop}%
\bibitem [{\citenamefont {Adare}\ \emph
  {et~al.}(2015{\natexlab{b}})\citenamefont {Adare} \emph
  {et~al.}}]{Adare:2014keg}%
  \BibitemOpen
  \bibfield  {author} {\bibinfo {author} {\bibfnamefont {A.}~\bibnamefont
  {Adare}} \emph {et~al.} (\bibinfo {collaboration} {PHENIX Collaboration}),\
  }\bibfield  {title} {\bibinfo {title} {{Measurement of long-range angular
  correlation and quadrupole anisotropy of pions and (anti)protons in central
  $d+$Au collisions at $\sqrt{s_{_{NN}}}$=200 GeV}},\ }\href
  {https://doi.org/10.1103/PhysRevLett.114.192301} {\bibfield  {journal}
  {\bibinfo  {journal} {Phys. Rev. Lett.}\ }\textbf {\bibinfo {volume} {114}},\
  \bibinfo {pages} {192301} (\bibinfo {year} {2015}{\natexlab{b}})}\BibitemShut
  {NoStop}%
\bibitem [{\citenamefont {Adare}\ \emph
  {et~al.}(2013{\natexlab{b}})\citenamefont {Adare} \emph
  {et~al.}}]{Adare:2013piz}%
  \BibitemOpen
  \bibfield  {author} {\bibinfo {author} {\bibfnamefont {A.}~\bibnamefont
  {Adare}} \emph {et~al.} (\bibinfo {collaboration} {PHENIX Collaboration}),\
  }\bibfield  {title} {\bibinfo {title} {{Quadrupole Anisotropy in Dihadron
  Azimuthal Correlations in Central $d+$Au Collisions at $\sqrt{s_{_{NN}}}$=200
  GeV}},\ }\href {https://doi.org/10.1103/PhysRevLett.111.212301} {\bibfield
  {journal} {\bibinfo  {journal} {Phys. Rev. Lett.}\ }\textbf {\bibinfo
  {volume} {111}},\ \bibinfo {pages} {212301} (\bibinfo {year}
  {2013}{\natexlab{b}})}\BibitemShut {NoStop}%
\bibitem [{\citenamefont {Adare}\ \emph
  {et~al.}(2018{\natexlab{b}})\citenamefont {Adare} \emph
  {et~al.}}]{Adare:2017rdq}%
  \BibitemOpen
  \bibfield  {author} {\bibinfo {author} {\bibfnamefont {A.}~\bibnamefont
  {Adare}} \emph {et~al.} (\bibinfo {collaboration} {PHENIX Collaboration}),\
  }\bibfield  {title} {\bibinfo {title} {{Measurement of emission angle
  anisotropy via long-range angular correlations with high $p_T$ hadrons in
  $d+$Au and $p+p$ collisions at $\sqrt{s_{_{NN}}}=200$ GeV}},\ }\href
  {https://doi.org/10.1103/PhysRevC.98.014912} {\bibfield  {journal} {\bibinfo
  {journal} {Phys. Rev. C}\ }\textbf {\bibinfo {volume} {98}},\ \bibinfo
  {pages} {014912} (\bibinfo {year} {2018}{\natexlab{b}})}\BibitemShut
  {NoStop}%
\bibitem [{\citenamefont {Adare}\ \emph
  {et~al.}(2018{\natexlab{c}})\citenamefont {Adare} \emph
  {et~al.}}]{Adare:2017wlc}%
  \BibitemOpen
  \bibfield  {author} {\bibinfo {author} {\bibfnamefont {A.}~\bibnamefont
  {Adare}} \emph {et~al.} (\bibinfo {collaboration} {PHENIX Collaboration}),\
  }\bibfield  {title} {\bibinfo {title} {{Measurements of mass-dependent
  azimuthal anisotropy in central $p+$Au, $d+$Au, and $^3$He$+$Au collisions at
  $\sqrt{s_{_{NN}}}=200$ GeV}},\ }\href
  {https://doi.org/10.1103/PhysRevC.97.064904} {\bibfield  {journal} {\bibinfo
  {journal} {Phys. Rev. C}\ }\textbf {\bibinfo {volume} {97}},\ \bibinfo
  {pages} {064904} (\bibinfo {year} {2018}{\natexlab{c}})}\BibitemShut
  {NoStop}%
\bibitem [{\citenamefont {Aidala}\ \emph {et~al.}(2019)\citenamefont {Aidala}
  \emph {et~al.}}]{PHENIX:2018lia}%
  \BibitemOpen
  \bibfield  {author} {\bibinfo {author} {\bibfnamefont {C.}~\bibnamefont
  {Aidala}} \emph {et~al.} (\bibinfo {collaboration} {PHENIX Collaboration}),\
  }\bibfield  {title} {\bibinfo {title} {{Creation of quark-gluon plasma
  droplets with three distinct geometries}},\ }\href
  {https://doi.org/10.1038/s41567-018-0360-0} {\bibfield  {journal} {\bibinfo
  {journal} {Nature Phys.}\ }\textbf {\bibinfo {volume} {15}},\ \bibinfo
  {pages} {214} (\bibinfo {year} {2019})}\BibitemShut {NoStop}%
\bibitem [{\citenamefont {Shen}\ \emph {et~al.}(2017)\citenamefont {Shen},
  \citenamefont {Paquet}, \citenamefont {Denicol}, \citenamefont {Jeon},\ and\
  \citenamefont {Gale}}]{PhysRevC.95.014906}%
  \BibitemOpen
  \bibfield  {author} {\bibinfo {author} {\bibfnamefont {C.}~\bibnamefont
  {Shen}}, \bibinfo {author} {\bibfnamefont {J.-F.}\ \bibnamefont {Paquet}},
  \bibinfo {author} {\bibfnamefont {G.~S.}\ \bibnamefont {Denicol}}, \bibinfo
  {author} {\bibfnamefont {S.}~\bibnamefont {Jeon}},\ and\ \bibinfo {author}
  {\bibfnamefont {C.}~\bibnamefont {Gale}},\ }\bibfield  {title} {\bibinfo
  {title} {{Collectivity and electromagnetic radiation in small systems}},\
  }\href {https://doi.org/10.1103/PhysRevC.95.014906} {\bibfield  {journal}
  {\bibinfo  {journal} {Phys. Rev. C}\ }\textbf {\bibinfo {volume} {95}},\
  \bibinfo {pages} {014906} (\bibinfo {year} {2017})}\BibitemShut {NoStop}%
\bibitem [{\citenamefont {Mace}\ \emph {et~al.}(2018)\citenamefont {Mace},
  \citenamefont {Skokov}, \citenamefont {Tribedy},\ and\ \citenamefont
  {Venugopalan}}]{Mace:2018vwq}%
  \BibitemOpen
  \bibfield  {author} {\bibinfo {author} {\bibfnamefont {M.}~\bibnamefont
  {Mace}}, \bibinfo {author} {\bibfnamefont {V.~V.}\ \bibnamefont {Skokov}},
  \bibinfo {author} {\bibfnamefont {P.}~\bibnamefont {Tribedy}},\ and\ \bibinfo
  {author} {\bibfnamefont {R.}~\bibnamefont {Venugopalan}},\ }\bibfield
  {title} {\bibinfo {title} {{Hierarchy of Azimuthal Anisotropy Harmonics in
  Collisions of Small Systems from the Color Glass Condensate}},\ }\href
  {https://doi.org/10.1103/PhysRevLett.121.052301} {\bibfield  {journal}
  {\bibinfo  {journal} {Phys. Rev. Lett.}\ }\textbf {\bibinfo {volume} {121}},\
  \bibinfo {pages} {052301} (\bibinfo {year} {2018})},\ \bibinfo {note}
  {[Erratum: Phys.Rev.Lett. 123, 039901 (2019)]}\BibitemShut {NoStop}%
\bibitem [{\citenamefont {Mace}\ \emph {et~al.}(2019)\citenamefont {Mace},
  \citenamefont {Skokov}, \citenamefont {Tribedy},\ and\ \citenamefont
  {Venugopalan}}]{Mace:2018yvl}%
  \BibitemOpen
  \bibfield  {author} {\bibinfo {author} {\bibfnamefont {M.}~\bibnamefont
  {Mace}}, \bibinfo {author} {\bibfnamefont {V.~V.}\ \bibnamefont {Skokov}},
  \bibinfo {author} {\bibfnamefont {P.}~\bibnamefont {Tribedy}},\ and\ \bibinfo
  {author} {\bibfnamefont {R.}~\bibnamefont {Venugopalan}},\ }\bibfield
  {title} {\bibinfo {title} {{Systematics of azimuthal anisotropy harmonics in
  proton-nucleus collisions at the LHC from the Color Glass Condensate}},\
  }\href {https://doi.org/10.1016/j.physletb.2018.09.064} {\bibfield  {journal}
  {\bibinfo  {journal} {Phys. Lett. B}\ }\textbf {\bibinfo {volume} {788}},\
  \bibinfo {pages} {161} (\bibinfo {year} {2019})},\ \bibinfo {note} {[Erratum:
  Phys.Lett.B 799, 135006 (2019)]}\BibitemShut {NoStop}%
\bibitem [{\citenamefont {Nagle}\ and\ \citenamefont
  {Zajc}(2019)}]{Nagle:2018ybc}%
  \BibitemOpen
  \bibfield  {author} {\bibinfo {author} {\bibfnamefont {J.~L.}\ \bibnamefont
  {Nagle}}\ and\ \bibinfo {author} {\bibfnamefont {W.~A.}\ \bibnamefont
  {Zajc}},\ }\bibfield  {title} {\bibinfo {title} {{Assessing saturation
  physics explanations of collectivity in small collision systems with the
  IP-Jazma model}},\ }\href {https://doi.org/10.1103/PhysRevC.99.054908}
  {\bibfield  {journal} {\bibinfo  {journal} {Phys. Rev. C}\ }\textbf {\bibinfo
  {volume} {99}},\ \bibinfo {pages} {054908} (\bibinfo {year}
  {2019})}\BibitemShut {NoStop}%
\bibitem [{\citenamefont {Schenke}\ \emph {et~al.}(2020)\citenamefont
  {Schenke}, \citenamefont {Shen},\ and\ \citenamefont
  {Tribedy}}]{Schenke:2019pmk}%
  \BibitemOpen
  \bibfield  {author} {\bibinfo {author} {\bibfnamefont {B.}~\bibnamefont
  {Schenke}}, \bibinfo {author} {\bibfnamefont {C.}~\bibnamefont {Shen}},\ and\
  \bibinfo {author} {\bibfnamefont {P.}~\bibnamefont {Tribedy}},\ }\bibfield
  {title} {\bibinfo {title} {{Hybrid Color Glass Condensate and hydrodynamic
  description of the Relativistic Heavy Ion Collider small system scan}},\
  }\href {https://doi.org/10.1016/j.physletb.2020.135322} {\bibfield  {journal}
  {\bibinfo  {journal} {Phys. Lett. B}\ }\textbf {\bibinfo {volume} {803}},\
  \bibinfo {pages} {135322} (\bibinfo {year} {2020})}\BibitemShut {NoStop}%
\bibitem [{\citenamefont {Romatschke}(2015)}]{Romatschke:2015gxa}%
  \BibitemOpen
  \bibfield  {author} {\bibinfo {author} {\bibfnamefont {P.}~\bibnamefont
  {Romatschke}},\ }\bibfield  {title} {\bibinfo {title} {{Light-Heavy Ion
  Collisions: A window into pre-equilibrium QCD dynamics?}},\ }\href
  {https://doi.org/10.1140/epjc/s10052-015-3509-3} {\bibfield  {journal}
  {\bibinfo  {journal} {Eur. Phys. J. C}\ }\textbf {\bibinfo {volume} {75}},\
  \bibinfo {pages} {305} (\bibinfo {year} {2015})}\BibitemShut {NoStop}%
\bibitem [{\citenamefont {Orjuela~Koop}\ \emph {et~al.}(2015)\citenamefont
  {Orjuela~Koop}, \citenamefont {Adare}, \citenamefont {McGlinchey},\ and\
  \citenamefont {Nagle}}]{Koop:2015wea}%
  \BibitemOpen
  \bibfield  {author} {\bibinfo {author} {\bibfnamefont {J.~D.}\ \bibnamefont
  {Orjuela~Koop}}, \bibinfo {author} {\bibfnamefont {A.}~\bibnamefont {Adare}},
  \bibinfo {author} {\bibfnamefont {D.}~\bibnamefont {McGlinchey}},\ and\
  \bibinfo {author} {\bibfnamefont {J.~L.}\ \bibnamefont {Nagle}},\ }\bibfield
  {title} {\bibinfo {title} {{Azimuthal anisotropy relative to the participant
  plane from a multiphase transport model in central p + Au , d + Au , and
  $^{3}$He + Au collisions at $\sqrt{s_{NN}}=200$ GeV}},\ }\href
  {https://doi.org/10.1103/PhysRevC.92.054903} {\bibfield  {journal} {\bibinfo
  {journal} {Phys. Rev. C}\ }\textbf {\bibinfo {volume} {92}},\ \bibinfo
  {pages} {054903} (\bibinfo {year} {2015})}\BibitemShut {NoStop}%
\bibitem [{\citenamefont {Welsh}\ \emph {et~al.}(2016)\citenamefont {Welsh},
  \citenamefont {Singer},\ and\ \citenamefont {Heinz}}]{Welsh:2016siu}%
  \BibitemOpen
  \bibfield  {author} {\bibinfo {author} {\bibfnamefont {K.}~\bibnamefont
  {Welsh}}, \bibinfo {author} {\bibfnamefont {J.}~\bibnamefont {Singer}},\ and\
  \bibinfo {author} {\bibfnamefont {U.~W.}\ \bibnamefont {Heinz}},\ }\bibfield
  {title} {\bibinfo {title} {{Initial state fluctuations in collisions between
  light and heavy ions}},\ }\href {https://doi.org/10.1103/PhysRevC.94.024919}
  {\bibfield  {journal} {\bibinfo  {journal} {Phys. Rev. C}\ }\textbf {\bibinfo
  {volume} {94}},\ \bibinfo {pages} {024919} (\bibinfo {year}
  {2016})}\BibitemShut {NoStop}%
\bibitem [{\citenamefont {Schenke}\ \emph {et~al.}(2012)\citenamefont
  {Schenke}, \citenamefont {Tribedy},\ and\ \citenamefont
  {Venugopalan}}]{Schenke:2012wb}%
  \BibitemOpen
  \bibfield  {author} {\bibinfo {author} {\bibfnamefont {B.}~\bibnamefont
  {Schenke}}, \bibinfo {author} {\bibfnamefont {P.}~\bibnamefont {Tribedy}},\
  and\ \bibinfo {author} {\bibfnamefont {R.}~\bibnamefont {Venugopalan}},\
  }\bibfield  {title} {\bibinfo {title} {{Fluctuating Glasma initial conditions
  and flow in heavy ion collisions}},\ }\href
  {https://doi.org/10.1103/PhysRevLett.108.252301} {\bibfield  {journal}
  {\bibinfo  {journal} {Phys. Rev. Lett.}\ }\textbf {\bibinfo {volume} {108}},\
  \bibinfo {pages} {252301} (\bibinfo {year} {2012})}\BibitemShut {NoStop}%
\bibitem [{\citenamefont {Adcox}\ \emph {et~al.}(2003)\citenamefont {Adcox}
  \emph {et~al.}}]{Adcox:2003zm}%
  \BibitemOpen
  \bibfield  {author} {\bibinfo {author} {\bibfnamefont {K.}~\bibnamefont
  {Adcox}} \emph {et~al.} (\bibinfo {collaboration} {PHENIX Collaboration}),\
  }\bibfield  {title} {\bibinfo {title} {{PHENIX detector overview}},\ }\href
  {https://doi.org/10.1016/S0168-9002(02)01950-2} {\bibfield  {journal}
  {\bibinfo  {journal} {Nucl. Instrum. Methods Phys. Res., Sec. A}\ }\textbf
  {\bibinfo {volume} {499}},\ \bibinfo {pages} {469} (\bibinfo {year}
  {2003})}\BibitemShut {NoStop}%
\bibitem [{\citenamefont {Akikawa}\ \emph {et~al.}(2003)\citenamefont {Akikawa}
  \emph {et~al.}}]{Akikawa:2003zs}%
  \BibitemOpen
  \bibfield  {author} {\bibinfo {author} {\bibfnamefont {H.}~\bibnamefont
  {Akikawa}} \emph {et~al.} (\bibinfo {collaboration} {PHENIX Collaboration}),\
  }\bibfield  {title} {\bibinfo {title} {{PHENIX muon arms}},\ }\href
  {https://doi.org/10.1016/S0168-9002(02)01955-1} {\bibfield  {journal}
  {\bibinfo  {journal} {Nucl. Instrum. Methods Phys. Res., Sec. A}\ }\textbf
  {\bibinfo {volume} {499}},\ \bibinfo {pages} {537} (\bibinfo {year}
  {2003})}\BibitemShut {NoStop}%
\bibitem [{\citenamefont {Allen}\ \emph {et~al.}(2003)\citenamefont {Allen}
  \emph {et~al.}}]{ALLEN2003549}%
  \BibitemOpen
  \bibfield  {author} {\bibinfo {author} {\bibfnamefont {M.}~\bibnamefont
  {Allen}} \emph {et~al.},\ }\bibfield  {title} {\bibinfo {title} {{PHENIX
  inner detectors}},\ }\href
  {https://doi.org/https://doi.org/10.1016/S0168-9002(02)01956-3} {\bibfield
  {journal} {\bibinfo  {journal} {Nucl. Instrum. Methods Phys. Res., Sec. A}\
  }\textbf {\bibinfo {volume} {499}},\ \bibinfo {pages} {549} (\bibinfo {year}
  {2003})},\ \bibinfo {note} {the Relativistic Heavy Ion Collider Project: RHIC
  and its Detectors}\BibitemShut {NoStop}%
\bibitem [{\citenamefont {Aidala}\ \emph {et~al.}(2014)\citenamefont {Aidala}
  \emph {et~al.}}]{Aidala:2013vna}%
  \BibitemOpen
  \bibfield  {author} {\bibinfo {author} {\bibfnamefont {C.}~\bibnamefont
  {Aidala}} \emph {et~al.},\ }\bibfield  {title} {\bibinfo {title} {{The PHENIX
  Forward Silicon Vertex Detector}},\ }\href
  {https://doi.org/10.1016/j.nima.2014.04.017} {\bibfield  {journal} {\bibinfo
  {journal} {Nucl. Instrum. Methods Phys. Res., Sec. A}\ }\textbf {\bibinfo
  {volume} {755}},\ \bibinfo {pages} {44} (\bibinfo {year} {2014})}\BibitemShut
  {NoStop}%
\bibitem [{\citenamefont {Ollitrault}\ \emph {et~al.}(2009)\citenamefont
  {Ollitrault}, \citenamefont {Poskanzer},\ and\ \citenamefont
  {Voloshin}}]{Ollitrault:2009ie}%
  \BibitemOpen
  \bibfield  {author} {\bibinfo {author} {\bibfnamefont {J.-Y.}\ \bibnamefont
  {Ollitrault}}, \bibinfo {author} {\bibfnamefont {A.~M.}\ \bibnamefont
  {Poskanzer}},\ and\ \bibinfo {author} {\bibfnamefont {S.~A.}\ \bibnamefont
  {Voloshin}},\ }\bibfield  {title} {\bibinfo {title} {{Effect of flow
  fluctuations and nonflow on elliptic flow methods}},\ }\href
  {https://doi.org/10.1103/PhysRevC.80.014904} {\bibfield  {journal} {\bibinfo
  {journal} {Phys. Rev. C}\ }\textbf {\bibinfo {volume} {80}},\ \bibinfo
  {pages} {014904} (\bibinfo {year} {2009})}\BibitemShut {NoStop}%
\bibitem [{\citenamefont {Poskanzer}\ and\ \citenamefont
  {Voloshin}(1998)}]{Poskanzer:1998yz}%
  \BibitemOpen
  \bibfield  {author} {\bibinfo {author} {\bibfnamefont {A.~M.}\ \bibnamefont
  {Poskanzer}}\ and\ \bibinfo {author} {\bibfnamefont {S.~A.}\ \bibnamefont
  {Voloshin}},\ }\bibfield  {title} {\bibinfo {title} {{Methods for analyzing
  anisotropic flow in relativistic nuclear collisions}},\ }\href
  {https://doi.org/10.1103/PhysRevC.58.1671} {\bibfield  {journal} {\bibinfo
  {journal} {Phys. Rev. C}\ }\textbf {\bibinfo {volume} {58}},\ \bibinfo
  {pages} {1671} (\bibinfo {year} {1998})}\BibitemShut {NoStop}%
\bibitem [{sup()}]{supp_mat}%
  \BibitemOpen
  \href@noop {} {}\bibinfo {note} {{See Supplemental Material at [URL will be
  inserted by publisher] for a set of example two-particle correlations and a
  complete set of extracted Fourier coefficients ($c_1$, $c_2$, $c_3$, $c_4$)
  and their statistical uncertainties}}\BibitemShut {NoStop}%
\bibitem [{\citenamefont {Lim}\ \emph {et~al.}(2019)\citenamefont {Lim},
  \citenamefont {Hu}, \citenamefont {Belmont}, \citenamefont {Hill},
  \citenamefont {Nagle},\ and\ \citenamefont {Perepelitsa}}]{Lim:2019cys}%
  \BibitemOpen
  \bibfield  {author} {\bibinfo {author} {\bibfnamefont {S.~H.}\ \bibnamefont
  {Lim}}, \bibinfo {author} {\bibfnamefont {Q.}~\bibnamefont {Hu}}, \bibinfo
  {author} {\bibfnamefont {R.}~\bibnamefont {Belmont}}, \bibinfo {author}
  {\bibfnamefont {K.~K.}\ \bibnamefont {Hill}}, \bibinfo {author}
  {\bibfnamefont {J.~L.}\ \bibnamefont {Nagle}},\ and\ \bibinfo {author}
  {\bibfnamefont {D.~V.}\ \bibnamefont {Perepelitsa}},\ }\bibfield  {title}
  {\bibinfo {title} {{Examination of flow and nonflow factorization methods in
  small collision systems}},\ }\href
  {https://doi.org/10.1103/PhysRevC.100.024908} {\bibfield  {journal} {\bibinfo
   {journal} {Phys. Rev. C}\ }\textbf {\bibinfo {volume} {100}},\ \bibinfo
  {pages} {024908} (\bibinfo {year} {2019})}\BibitemShut {NoStop}%
\end{thebibliography}

%
 
\end{document}